\documentclass[aoas,preprint]{imsart}

\RequirePackage[OT1]{fontenc}
\RequirePackage{amsthm,amsmath}
\RequirePackage{natbib}
\RequirePackage[colorlinks,citecolor=blue,urlcolor=blue]{hyperref}
\usepackage{graphicx}
\usepackage{amsfonts}
\usepackage{subfig}
\usepackage[utf8]{inputenc}
\usepackage{multirow}
\usepackage{tikz}

\usepackage{xcolor}
\usepackage{mathtools}

\newcommand \contact{\phi}
\newcommand \noise{\epsilon}
\newcommand \common{w}
\newcommand \jesa{\mathbb{J}}
\newcommand \kernel{k}

\newcommand \pop{\mathbb{A}}

\arxiv{arXiv:0000.0000}

\startlocaldefs
\numberwithin{equation}{section}
\theoremstyle{plain}

\endlocaldefs

\begin{document}

\begin{frontmatter}
\title{A Bayesian Hierarchical Model for Evaluating Forensic Footwear Evidence}
\runtitle{Bayesian Footwear Analysis}

\begin{aug}
\author{\fnms{Neil A.} \snm{Spencer}\thanksref{t2}\ead[label=e1]{nspencer@andrew.cmu.edu}},
\and
\author{\fnms{Jared S.} \snm{Murray}\ead[label=e2]{jared.murray@mccombs.utexas.edu}}

\thankstext{t2}{Supported in part by the Center for Statistics and Applications in Forensic Evidence (CSAFE) through Cooperative Agreement \#70NANB15H176 between NIST and Iowa State University, which includes activities carried out at Carnegie Mellon University, University of California Irvine, and University of Virginia, as well as a fellowship from the Natural Science and Research Council of Canada.}
\runauthor{N. Spencer and J.S. Murray}

\affiliation{Carnegie Mellon University and University of Texas at Austin}

\address{132 Baker Hall\\
Carnegie Mellon University\\
Pittsburgh, PA, USA 15213\\
\printead{e1}\\
}

\address{2110 Speedway\\
B6500\\
Austin, Texas, USA 78712\\
\printead{e2}\\
}
\end{aug}

\begin{abstract}
When a latent shoeprint is discovered at a crime scene, forensic analysts inspect it for distinctive patterns of wear such as scratches and holes (known as accidentals) on the source shoe's sole. If its accidentals correspond to those of a suspect's shoe, the print can be used as forensic evidence to place the suspect at the crime scene. The strength of this evidence depends on the random match probability--- the chance that a shoe chosen at random would match the crime scene print's accidentals. Evaluating random match probabilities requires an accurate model for the spatial distribution of accidentals on shoe soles. A recent report by the President's Council of Advisors in Science and Technology criticized existing models in the literature, calling for new empirically validated techniques. We respond to this request with a new spatial point process model\footnote{Code and synthetic data is available at \url{github.com/neilspencer/cindRella/}.} for accidental locations, developed within a hierarchical Bayesian framework. We treat the tread pattern of each shoe as a covariate, allowing us to pool information across large heterogeneous databases of shoes. Existing models ignore this information; our results show that including it leads to significantly better model fit. We demonstrate this by fitting our model to one such database.
\end{abstract}



\end{frontmatter}

\section{Introduction}\label{intro}

\begin{figure}[h]
\centering
\subfloat[]{\includegraphics[height = 2in]{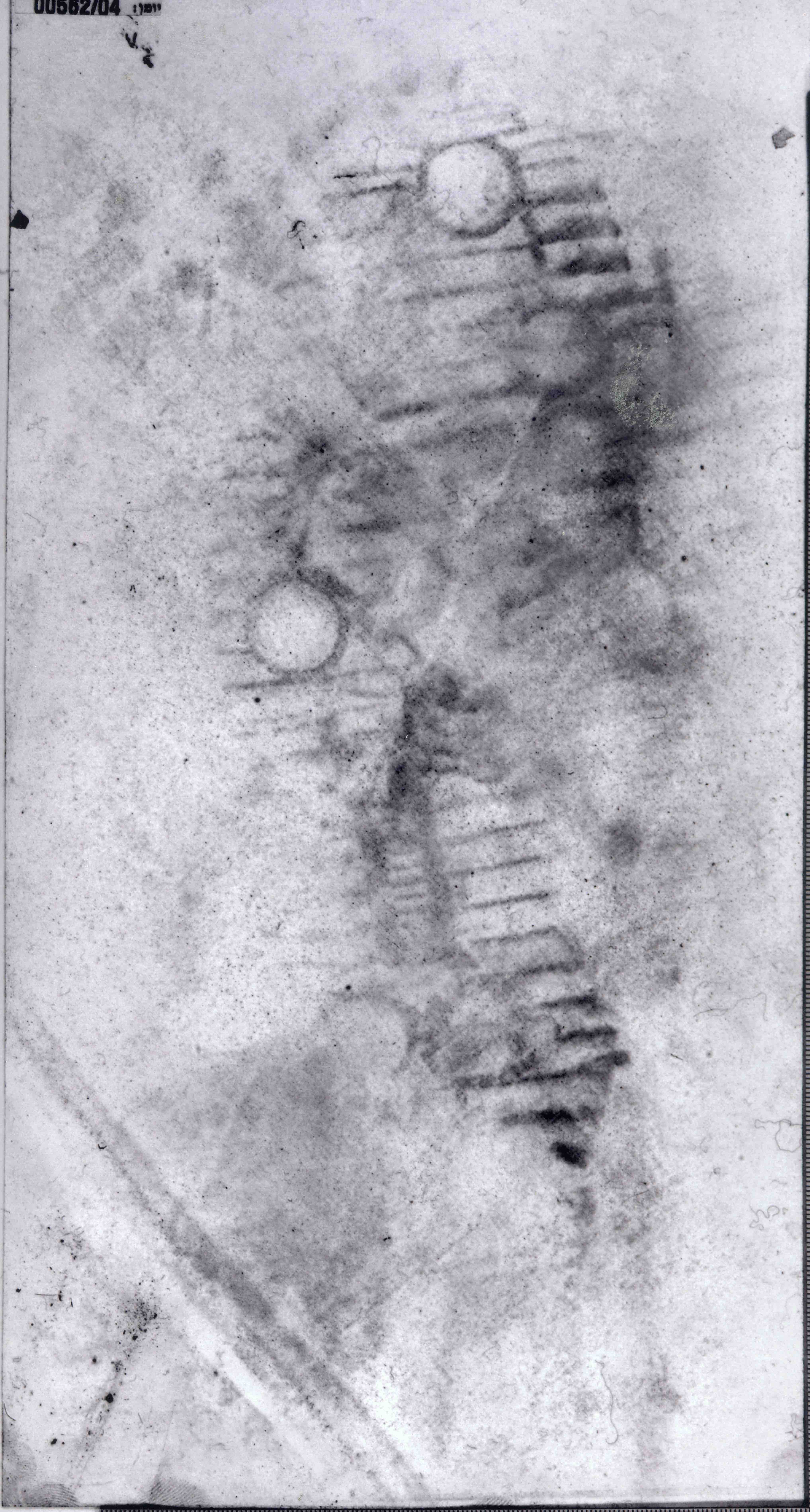}\label{latent}} \hspace{1cm}
\subfloat[]{\includegraphics[height = 2in]{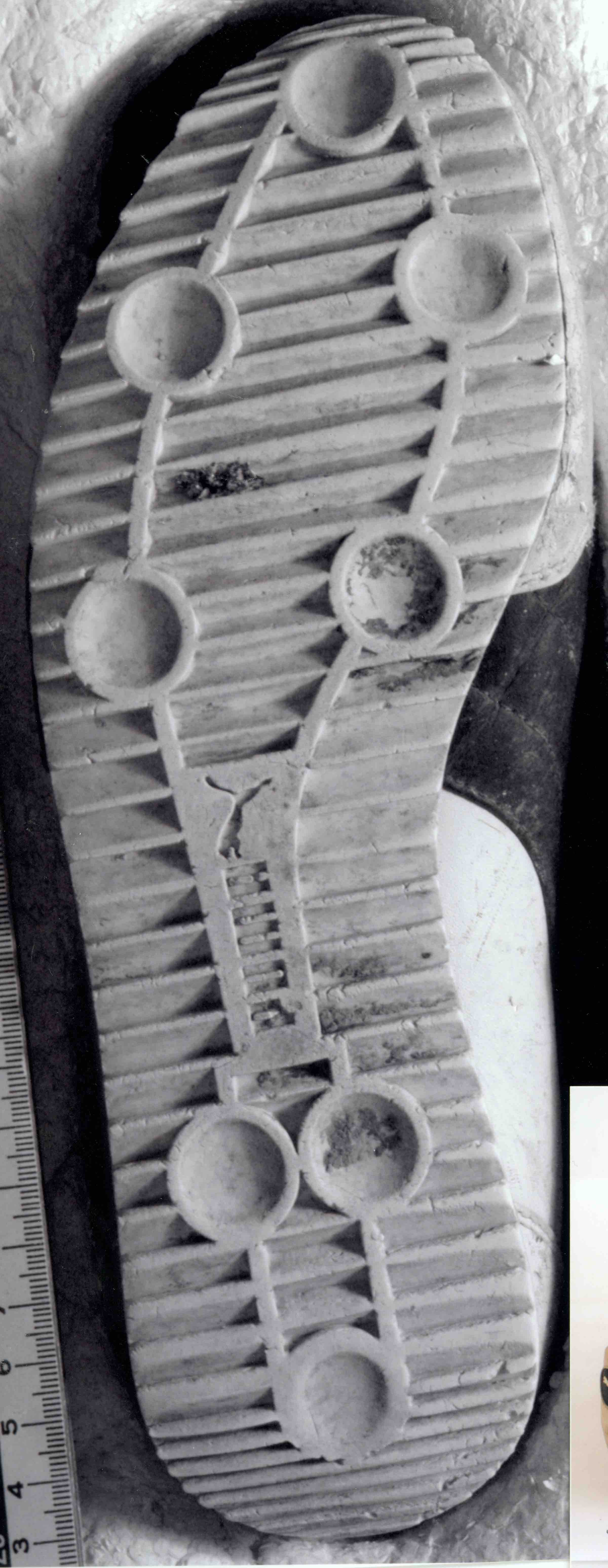}\label{shoe}} \hspace{1cm}
\subfloat[]{\includegraphics[height = 2in]{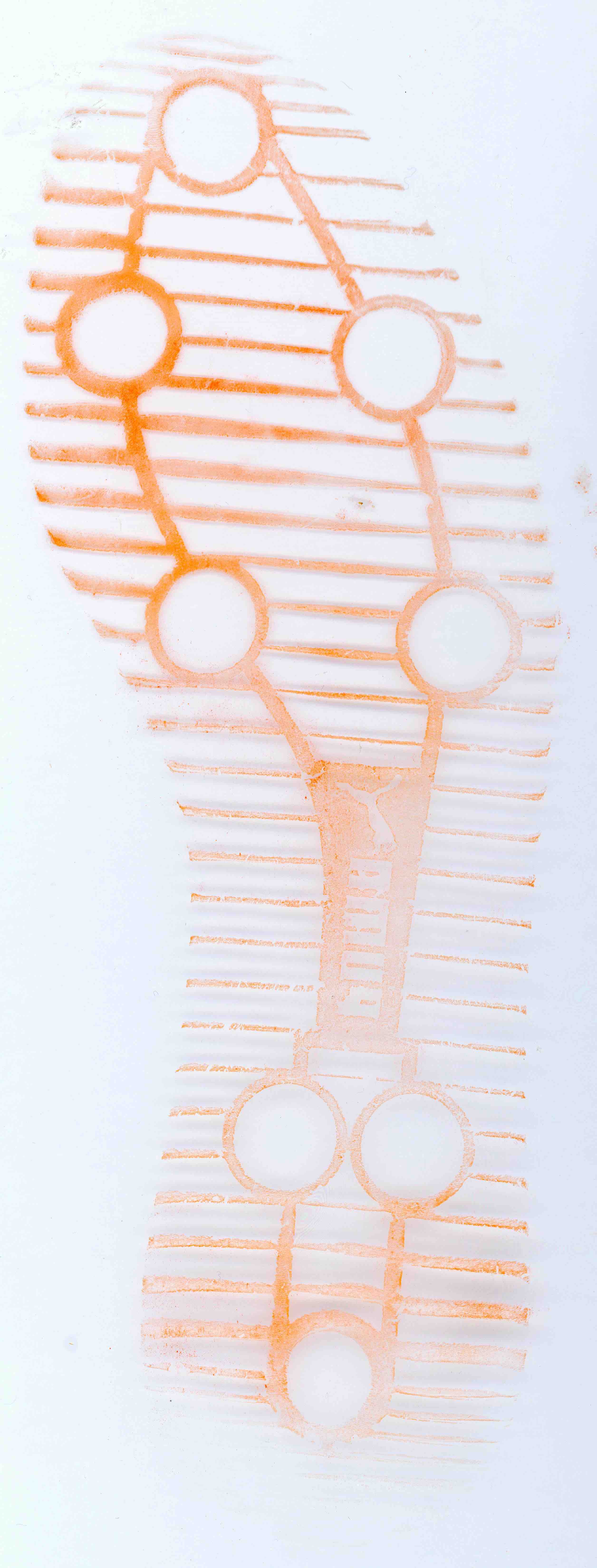}\label{print}} \hspace{1cm}
\subfloat[]{\includegraphics[height = 2in]{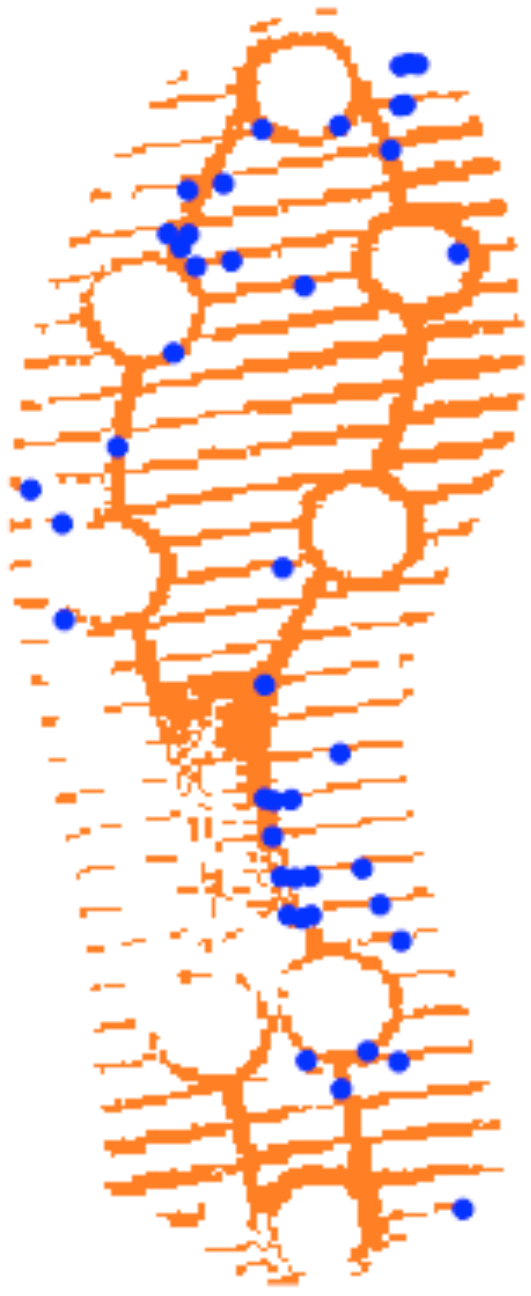}\label{accidentals}} 
\caption{(a)-(d) represent objects pertaining to the same shoe from the JESA database. (a) is a photograph of a latent crime scene print, (b) a photo of the shoe's sole, (c) is a raw image of a test impression, and (d) is the contact surface obtained from standardizing the test impression. The superimposed blue points in (d) correspond to accidental locations.}
\end{figure}

Forensic footwear analysis encompasses a suite of techniques used to analyze latent shoeprints as part of forensic investigations. A principal goal of these investigations is to link a suspect's shoe to a crime scene print, providing evidence to place the suspect at the scene of the crime. Figure~\ref{latent} provides an example of a latent crime scene shoeprint. 

As described by \cite{bodziak2017forensic}, the procedure for determining the source of a latent print typically consists of two stages. First, the examiner inspects the tread of the latent print to identify class characteristics (brand, model, and size) of the source shoe. This identification can be carried out manually, or automated using tread matching algorithms (e.g. \cite{srihari2014computational, richetelli2017classification, kong2017cross}).

Manufacturers routinely produce thousands of shoes of the same make and model, meaning that class characteristics alone are often insufficient for determining a print's source. For this reason examiners regularly turn to a second stage of analysis: the inspection of \emph{accidentals}. Accidentals, also known as randomly acquired characteristics, are the post-manufacturing cuts, scrapes, holes, and debris that accumulate on a shoe sole. Examiners are trained to identify accidentals on a shoe by inspecting both the shoe's sole and test impressions--- high quality prints created using the shoe in a controlled laboratory setting. Figure~\ref{shoe}, Figure~\ref{print}, and Figure~\ref{accidentals} depict a shoe sole, test impression, and  accidentals locations, respectively. These images all correspond to the same shoe obtained from the JESA database \citep{yekutieli2012expert} (we describe the JESA database in \S\ref{data}).

In theory, if both the class characteristics and the accidentals of a suspect's shoe coincide exactly with those detected from the crime scene print, then the suspect's shoe is almost certainly the source of the print. In practice, the comparison is less clear-cut. Latent crime scene prints are typically of low quality, making it difficult to pick out all of the individual accidentals. Furthermore, accidental locations are known to vary slightly from test print to test print due to variability in the impression-taking process \citep{shor2017inherent}, so there is some uncertainty on their exact locations on the source shoe. As a result, accidental comparisons typically involve comparing a subset of approximate accidental locations on the test impression to those detected on the crime scene print. This uncertainty leaves the possibility of a false positive due to chance, especially for partial prints and tread patterns on which accidentals are very likely to occur in certain regions. 

To account for the possibility of a false positive, shoeprint analysts are encouraged to provide a measure of the uncertainty of the match when testifying in court \citep{edwards2009strengthening}. One popular summary for communicating this uncertainty is the random match probability (RMP) \citep{thompson2015lay}. The RMP is the probability that a randomly sampled shoe would produce a print matching the observed features at the crime scene. For instance, if 15 out of 10000 relevant shoes were consistent with the crime scene print, the RMP would be 0.0015.


The standard approach for evaluating RMPs decomposes into three terms: the evidence given by the class characteristics, the evidence based on general wear, and accidental-based evidence \citep{evett1998bayesian, skerrett2011bayesian}. In this work, we focus on accidental-based evidence, inspired by the recent report on forensic science by The President's Council for Advisors on Science \citep{president2016report} that criticized existing work in the area.

We address the concerns of \citet{president2016report} by developing and estimating the parameters of a model for the distribution of accidental configurations on a shoe. Specifically, we model the spatial distribution of accidentals on a shoe sole as a point process, treating the sole's tread pattern as a covariate. We fit and evaluate our model using the \emph{JESA database} \citep{yekutieli2012expert}, a ground truth dataset of 386 accidental-annotated shoeprints compiled by the Israeli Police Department's Division of Forensic Science. The JESA database is one of the largest existing databases of its kind \citep{speir2016quantifying}, consisting of shoes with a variety of tread patterns. 

We define our model within a hierarchical Bayesian framework, pooling information across JESA to infer general trends spanning a broad variety of shoes. Our model is a finite resolution version of the normalized compound random measure framework of \cite{griffin2017compound}, modified to incorporate spatial covariates and dependency of the intensity across space. We develop the computational tools to fit our model, evaluate it, and demonstrate that it outperforms existing approaches by a wide margin.

The remainder of this paper is organized as follows. In Section~\ref{pre}, we review the literature related to random match probabilities, formalize the link between evaluating random match probabilities and modeling spatial distributions of accidentals, describe the JESA database of annotated shoeprints collected by \citet{yekutieli2012expert}, and review the relevant literature pertaining to vectors of dependent probability measures. In Section~\ref{model}, we provide the details of our hierarchical Bayesian model for spatial configurations of accidentals. In Section~\ref{computation}, we propose a Markov chain Monte Carlo algorithm for inferring the parameters of the model, and an importance sampling algorithm for evaluating marginal likelihoods. In Section~\ref{comparisons}, we showcase the results of fitting our model to the JESA dataset and compare its performance to other candidate models. Section~\ref{discussion} contains some concluding remarks.

\section{Preliminaries}\label{pre}

%


\subsection{Random Match Probabilities}\label{sec:rmp}

A theory to evaluate RMPs for footwear evidence was laid out in \cite{evett1998bayesian} in the context of evaluating likelihood ratios. The framework is equally applicable to evaluating raw RMPs. Let $y$ denote a crime scene print and $\pop$ denote the relevant population of plausible sources of the crime scene print. For instance, $\pop$ could be all shoes belonging to residents of a particular city or town. As per \cite{evett1998bayesian}, the random match probability for footwear evidence is given by 
\begin{align}
   \text{RMP} &= p( y \equiv s \mid s \sim \pop)
\end{align}
where $y \equiv s$ indicates that shoe $s$ exhibits features consistent with those of the print $y$, and $s \sim \pop$ is shorthand for $s$ being chosen uniformly at random from all shoes in the set $\pop$. Further discussion of random match probabilities with examples from forensic science is available in \cite{srihari2011generative}.

Following the classical two step process of forensic footwear analysis, \cite{evett1998bayesian} suggested that the RMP be calculated using the factorization $\text{RMP}= \text{rmp}_M  \text{rmp}_U$. Here, $\text{rmp}_M$ denotes the probability that a randomly chosen shoe in $\pop$ has class characteristics matching the latent crime scene print, and $\text{rmp}_U$ denotes the probability that the shoe is also consistent with the wear patterns and accidentals, given that it matches on the class characteristics. \cite{skerrett2011bayesian} refined this representation by further decomposing $\text{rmp}_U$ into $\text{rmp}_W$ and $\text{rmp}_V$, corresponding to separate conditional probabilities of matching on general wear and accidentals, respectively.

Let $y_M$, $y_W$, and $y_V$ denote the class characteristics, general wear, and accidentals observed on the latent print $y$ with $s_M$, $s_W$, $s_V$ denoting the same features as observed on a shoe $s \in \mathbb{A}$. The factorization proposed by \cite{skerrett2011bayesian} can be formally expressed as
\begin{align}
   \text{RMP} &= \text{rmp}_M \cdot \text{rmp}_W \cdot \text{rmp}_V,  \\\
           \text{rmp}_M   &= p(y_M \equiv s_M \mid s \sim \pop),\\
           \text{rmp}_W &= p(y_W \equiv s_W \mid s \sim \{s' \in \pop: y_M \equiv s'_M\}),\\
           \text{rmp}_V &= p(y_V \equiv s_V \mid  s \sim \{s' \in \pop: y_M \equiv s'_M, y_W \equiv s'_W\}) \label{rmp1},
\end{align}
where $y_M \equiv s_M$ denotes the class characteristics of $s$ being consistent with those of $y$, and $y_W \equiv s_W$ and $y_V \equiv s_V$ defined similarly. Implicit in this decomposition is the assumption that $y \equiv s$ is characterized by $y_M \equiv s_M$, $y_W \equiv s_W$, and $y_V \equiv s_V$, a reasonable choice given that these features form the basis of forensic footwear analysis \citep{bodziak2017forensic}. Strategies for evaluating $\text{rmp}_M$ and $\text{rmp}_W$ based on relevant databases (e.g. \citet{evett1998bayesian, champod2004establishing} for $\text{rmp}_M$ and \citet{fruchtenicht2002discrimination, facey1992shoe, bodziak2012determining} for $\text{rmp}_W$) were discussed in \cite{skerrett2011bayesian}. However, evaluating the accidental-based component $\text{rmp}_V$ was left as a subject for future work. In this work, we focus on the remaining accidental-based component. We begin by making two simplifying assumptions.

First, we follow \cite{petraco2010statistical} in assuming that the evidence present in a configuration of accidentals on a crime scene print $y_V$ is characterized by the locations (e.g. the blue points shown in Figure~\ref{accidentals}). We omit secondary characteristics such as shape or size of the accidental as they are difficult to reliably glean from latent prints. We use $x^s$ to denote the accidental locations on shoe $s$ and $x^y$ to denote the locations detected on print $y$. Employing a standardized coordinate system (details provided in \S \ref{data}), we have $x^s \in ([0,100] \times [0,200])^{N_s}$, $x^y \in ([0,100] \times [0,200])^{N_y}$ where $N_s$ denotes the number of accidentals on shoe $s$ and $N_y$ denotes the number detectable on print $y$. We use $x^s_n = (x^s_{n,1}, x^s_{n,2})$ to denote the $n$th row of $x^s$. 

Because examiners are adept at recovering $y_M$ and $y_W$ from a shoeprint $y$, our second assumption is that a shoe's class characteristics and wear are characterized by its \emph{contact surface}. A shoe's contact surface refers to the portion of its sole that typically touches the ground when worn --- the part responsible for leaving prints. An example contact surface is provided in Figure~\ref{accidentals}. We provide a more detailed definition of contact surface in \S\ref{contactsurface}.  Letting $\mathcal{C}^s$ denote the contact surface of shoe $s$, this assumption can be formalized as $s, s' \in \pop$, $\mathcal{C}^{s} = \mathcal{C}^{s'}$ if and only if $s_M \equiv s'_M$ and $s_W \equiv s'_W$. 

\begin{figure}[h]
\centering
\subfloat[]{\input{examplematch1.tex} \label{exampleprocess1}}
\subfloat[]{
\begin{tikzpicture}[x=1pt,y=1pt]
\definecolor{fillColor}{RGB}{255,255,255}
\path[use as bounding box,fill=fillColor,fill opacity=0.00] (0,0) rectangle (108.41,216.81);
\begin{scope}
\path[clip] (  0.00,  0.00) rectangle (108.41,216.81);
\definecolor{drawColor}{RGB}{255,255,255}
\definecolor{fillColor}{RGB}{255,255,255}

\path[draw=drawColor,line width= 0.9pt,line join=round,line cap=round,fill=fillColor] (  0.00,  0.00) rectangle (108.41,216.81);
\end{scope}
\begin{scope}
\path[clip] ( 27.84, 27.84) rectangle ( 99.41,207.81);
\definecolor{drawColor}{RGB}{255,255,255}
\definecolor{fillColor}{RGB}{255,255,255}

\path[draw=drawColor,line width= 0.9pt,line join=round,line cap=round,fill=fillColor] ( 27.84, 27.84) rectangle ( 99.40,207.81);
\node[inner sep=0pt,outer sep=0pt,anchor=south west,rotate=  0.00] at ( 36.62,  56.88) {
	\pgfimage[width= 44.24pt,height=128.43pt,interpolate=false]{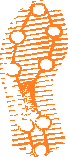}};
\definecolor{drawColor}{RGB}{0,0,255}
\definecolor{fillColor}{RGB}{0,0,255}

\path[draw=drawColor,line width= 0.4pt,line join=round,line cap=round,fill=fillColor] ( 64.89,173.52) circle (  1.96);

\path[draw=drawColor,line width= 0.4pt,line join=round,line cap=round,fill=fillColor] ( 70.10,175.53) circle (  1.96);

\path[draw=drawColor,line width= 0.4pt,line join=round,line cap=round,fill=fillColor] ( 70.14,179.54) circle (  1.96);

\path[draw=drawColor,line width= 0.4pt,line join=round,line cap=round,fill=fillColor] ( 70.84,179.77) circle (  1.96);

\path[draw=drawColor,line width= 0.4pt,line join=round,line cap=round,fill=fillColor] ( 71.67,179.69) circle (  1.96);

\path[draw=drawColor,line width= 0.4pt,line join=round,line cap=round,fill=fillColor] ( 51.91,162.53) circle (  1.96);

\path[draw=drawColor,line width= 0.4pt,line join=round,line cap=round,fill=fillColor] ( 52.55,159.32) circle (  1.96);

\path[draw=drawColor,line width= 0.4pt,line join=round,line cap=round,fill=fillColor] ( 58.24,173.14) circle (  1.96);

\path[draw=drawColor,line width= 0.4pt,line join=round,line cap=round,fill=fillColor] ( 59.88, 97.80) circle (  1.96);

\path[draw=drawColor,line width= 0.4pt,line join=round,line cap=round,fill=fillColor] ( 59.14,101.87) circle (  1.96);

\path[draw=drawColor,line width= 0.4pt,line join=round,line cap=round,fill=fillColor] ( 58.46,105.71) circle (  1.96);

\path[draw=drawColor,line width= 0.4pt,line join=round,line cap=round,fill=fillColor] ( 59.26,105.44) circle (  1.96);

\path[draw=drawColor,line width= 0.4pt,line join=round,line cap=round,fill=fillColor] ( 60.75,105.59) circle (  1.96);

\path[draw=drawColor,line width= 0.4pt,line join=round,line cap=round,fill=fillColor] ( 75.08,160.63) circle (  1.96);

\path[draw=drawColor,line width= 0.4pt,line join=round,line cap=round,fill=fillColor] ( 67.31, 80.22) circle (  1.96);

\path[draw=drawColor,line width= 0.4pt,line join=round,line cap=round,fill=fillColor] ( 62.05, 79.28) circle (  1.96);

\path[draw=drawColor,line width= 0.4pt,line join=round,line cap=round,fill=fillColor] ( 45.78,141.13) circle (  1.96);

\path[draw=drawColor,line width= 0.4pt,line join=round,line cap=round,fill=fillColor] ( 58.46,117.15) circle (  1.96);

\path[draw=drawColor,line width= 0.4pt,line join=round,line cap=round,fill=fillColor] ( 60.02,128.93) circle (  1.96);

\path[draw=drawColor,line width= 0.4pt,line join=round,line cap=round,fill=fillColor] ( 50.12,162.60) circle (  1.96);

\path[draw=drawColor,line width= 0.4pt,line join=round,line cap=round,fill=fillColor] ( 38.31,136.82) circle (  1.96);

\path[draw=drawColor,line width= 0.4pt,line join=round,line cap=round,fill=fillColor] ( 61.72, 93.61) circle (  1.96);

\path[draw=drawColor,line width= 0.4pt,line join=round,line cap=round,fill=fillColor] ( 69.30,171.06) circle (  1.96);

\path[draw=drawColor,line width= 0.4pt,line join=round,line cap=round,fill=fillColor] ( 41.22,123.69) circle (  1.96);

\path[draw=drawColor,line width= 0.4pt,line join=round,line cap=round,fill=fillColor] ( 60.49, 93.94) circle (  1.96);

\path[draw=drawColor,line width= 0.4pt,line join=round,line cap=round,fill=fillColor] ( 65.05, 76.42) circle (  1.96);

\path[draw=drawColor,line width= 0.4pt,line join=round,line cap=round,fill=fillColor] ( 75.53, 64.32) circle (  1.96);

\path[draw=drawColor,line width= 0.4pt,line join=round,line cap=round,fill=fillColor] ( 50.62,150.61) circle (  1.96);

\path[draw=drawColor,line width= 0.4pt,line join=round,line cap=round,fill=fillColor] ( 51.25,161.23) circle (  1.96);

\path[draw=drawColor,line width= 0.4pt,line join=round,line cap=round,fill=fillColor] ( 51.84,167.00) circle (  1.96);
\end{scope}
\end{tikzpicture} \label{exampleprocess2}} 
\subfloat[]{\input{examplematch3.tex} \label{exampleprocess3}} 
\caption{(a) depicts the accidental locations (blue) and contact surface (orange) for eight synthetic draws from the population $\pop_{\mathcal{C}^y}$ corresponding to the crime scene print $y$ shown in Figure~\ref{latent}. (b) depicts the contact surface $\mathcal{C}^y$ (orange) and accidental locations $x^y$ (blue). (c) illustrates the close correspondence between $x^y$ (blue) and $x^s$ (red) given by the accidental locations from the rectangle enclosed shoe in (a).}\label{exampleprocess}
\end{figure}

After characterizing $y_V$ using accidental locations and $y_W, w_M$ using the contact surface, we can now re-express the accidental-based random match probability in (\ref{rmp1}) in a form that is more tractable for statistical inference. The relation $y_V \equiv s_V$ reduces to a comparison of the point clouds $x^y$ and $x^s$ (denoted $x^y \equiv x^s$). The set $\{s' \in \pop: y_M \equiv s'_M, y_W \equiv s'_W\}$ reduces to the set of relevant shoes with the given contact surface (i.e. $\pop_{\mathcal{C}^y} = \{s' \in \pop: \mathcal{C}^{s'} = \mathcal{C}^y \}$, where $\mathcal{C}^y$ denotes the contact surface as determined from $y$). Thus, the accidental-based random match probability given in (\ref{rmp1}) reduces to
\begin{align}
           \text{rmp}_V &= p(x^y \equiv x^s \mid  s \sim \pop_{\mathcal{C}^y} )\label{rmpv}.
\end{align}
In theory, computing $\text{rmp}_V$ using (\ref{rmpv}) is straightforward. One would simply inspect all shoes in $\pop$ with contact surface $\mathcal{C}^y$ to determine the ratio that also have accidentals consistent with $x^y$. Even if $\pop$ were not completely accessible, a large random sample would suffice to provide a sufficiently accurate approximation. Figure~\ref{exampleprocess} illustrates this strategy for a small example.

In practice, the computation of $\text{rmp}_V$ is complicated by two issues:
\begin{enumerate}
\item In many cases, no shoes in $\pop_{\mathcal{C}^y}$ (other than the suspect's shoe) are accessible by the examiner. Examiners are left to rely on previous experience and limited data (e.g. a small convenience sample from $\pop$ or a related database) to make inferences regarding the conditional distribution of $x^s | s \sim \pop_{\mathcal{C}^y}$. Historically, these inferences have been based on heuristics that lack empirical support \citep{president2016report}. 

\item Determining if $x^y \equiv x^s$ is complicated by three phenomena: (i) a shoe's detected accidental locations are known to vary slightly each time it is printed \citep{shor2017inherent}, meaning that the locations in $x^y$ may only approximate those in $x^s$, (ii) some accidentals do not reliably show up on crime scene prints \citep{richetelli2017quantitative}, meaning that the accidentals in $x^y$ could be a thinned version of $x^s$, and (iii) test impressions may not be obtained until long after the crime was committed, leaving the opportunity for new accidentals to arise \citep{wyatt2005aging} or existing accidentals to change \citep{sheets2013shape} in the meantime.
\end{enumerate}

We concentrate on issue 1 in this paper, developing a more principled approach to inferring the distribution $x^s | s \sim \pop_{\mathcal{C}^y}$ using the JESA database. Issue 2 is beyond the scope of this paper, as determining an appropriate definition of $x^y \equiv x^s$ would require much richer data than is currently available in the literature. However, given a definition of $x^y \equiv x^s$, our model can compute the RMP via Monte Carlo. Figure~\ref{exampleprocess} demonstrates this process with Figure~\ref{exampleprocess1} depicting the samples drawn from the distribution $x^s | s \sim \pop_{\mathcal{C}^y}$. 
 
\subsection{JESA} \label{data}
The Jerusalem Shoeprint Accidentals Database (JESA) is one of a series of datasets created by the Israel Police Department's Division of Forensic Science. It pertains to 386 men's shoes collected as evidence through casework. A full description of the database is available in \cite{yekutieli2012expert}. For each shoe, there are two data structures relevant to our work -- the \emph{standardized shoeprint image} (contact surface) and the \emph{accidentals}.

\subsubsection{Standardized Shoeprint Image}\label{contactsurface}
Test impressions for each shoe were obtained by applying orange powder to their soles, pressing them onto clear films, then digitally photographing the residual orange impressions on the films. An example impression image is shown as Figure~\ref{print}. 

For consistency across shoes, each image was standardized onto a 200 by 100 grid. Standardization involved translating, aligning, and scaling the images so the prints were centered, pointed upwards, and of the same length. The axes for the alignment were designated through point-and-click software by trained examiners. All left shoes were mirrored to appear as right shoes. Alignment of the images facilitates the pooling of information across shoes, even if they differ in size or chirality (i.e. left shoe or right shoe).

After standardization, the images were smoothed and de-noised to isolate the \emph{contact surface}--- the areas of the shoe sole that typically touch the ground. The smoothing was performed to preserve the shoe's tread pattern and general wear while filtering out small breaks due to accidentals or imperfections in the impression. These contact surfaces take the form of 200 by 100 binary arrays, with each bit defining contact or non-contact of a region of the shoe. Figure \ref{accidentals} illustrates the positive values of contact surface for the shoe in Figure \ref{shoe}. The superimposed points are the locations of accidentals. 

\begin{figure}[h]
\centering
\subfloat[]{\includegraphics[height = 2in]{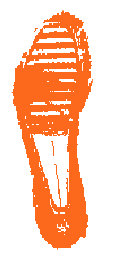}\label{contact1}} \hspace{1cm}
\subfloat[]{\includegraphics[height = 2in]{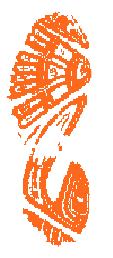}\label{contact2}} \hspace{1cm}
\subfloat[]{\includegraphics[height = 2in]{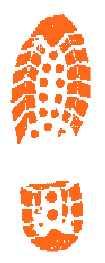}\label{contact3}}
\hspace{1cm}
\subfloat[]{\includegraphics[height = 2in]{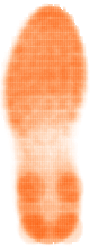}\label{aggregatedcontacts}}
\caption{(a),(b),(c) are standardized test impressions from the JESA database, (d) is the mean  test impression across the entire JESA database.}
\end{figure}

Additional example contact surfaces are shown in Figures~\ref{contact1}, \ref{contact2}, and \ref{contact3}, demonstrating the variety of tread patterns in the JESA database. No two contact surfaces in the JESA database are exactly alike, although those that correspond to the same brand of shoe are similar (differences in wear patterns, as well as variation in test impressions, account for the differences).

Figure~\ref{aggregatedcontacts} depicts the average contact surface across the entire database. It shows that it is far more common for regions of the shoe corresponding the heel and toes to be part of the contact surface than regions corresponding to the shoe arch. This discrepancy drives home the importance of conditioning on contact surface when evaluating accidental-based RMPs; shoes with arches that do make contact with the ground (the minority) would likely have different accidental distributions than those that do not. We use $\mathbb{C} = \left\{0, 1\right\}^{100 \times 200}$ to denote the space of values that a contact surface can take, and $\mathcal{C}^{s}\in \mathbb{C}$ to denote the contact surface of shoe $s$. 

\subsubsection{Accidentals} \label{data_accidentals}

For each shoe, examiners identified accidentals by inspecting the shoeprint image and the shoe sole itself. The locations of the centroids of the accidentals were recorded using a computerized system. These locations were stored as real numbers in $[0,100] \times [0,200]$ corresponding to the standardized space of the contact surface. The region $[0,1] \times [0,1]$ corresponds to the bottom left hand corner of the standardized grid, and $[99,100] \times [199, 200]$ corresponds to the top right. Figure~\ref{accidentals} gives an example of the locations accidentals as points on the shoeprint image. 

\begin{figure}[h]
\centering
\subfloat[]{
\begin{tikzpicture}[x=1pt,y=1pt]
\definecolor{fillColor}{RGB}{255,255,255}
\path[use as bounding box,fill=fillColor,fill opacity=0.00] (0,0) rectangle (144.54,144.54);
\begin{scope}
\path[clip] (  0.00,  0.00) rectangle (144.54,144.54);
\definecolor{drawColor}{RGB}{255,255,255}
\definecolor{fillColor}{RGB}{255,255,255}

\path[draw=drawColor,line width= 0.6pt,line join=round,line cap=round,fill=fillColor] (  0.00, -0.00) rectangle (144.54,144.54);
\end{scope}
\begin{scope}
\path[clip] ( 35.92, 30.72) rectangle (139.04,139.04);
\definecolor{fillColor}{RGB}{255,255,255}

\path[fill=fillColor] ( 35.92, 30.72) rectangle (139.04,139.04);
\definecolor{drawColor}{gray}{0.92}

\path[draw=drawColor,line width= 0.3pt,line join=round] ( 35.92, 47.72) --
	(139.04, 47.72);

\path[draw=drawColor,line width= 0.3pt,line join=round] ( 35.92, 71.85) --
	(139.04, 71.85);

\path[draw=drawColor,line width= 0.3pt,line join=round] ( 35.92, 95.98) --
	(139.04, 95.98);

\path[draw=drawColor,line width= 0.3pt,line join=round] ( 35.92,120.12) --
	(139.04,120.12);

\path[draw=drawColor,line width= 0.3pt,line join=round] ( 59.14, 30.72) --
	( 59.14,139.04);

\path[draw=drawColor,line width= 0.3pt,line join=round] ( 93.08, 30.72) --
	( 93.08,139.04);

\path[draw=drawColor,line width= 0.3pt,line join=round] (127.02, 30.72) --
	(127.02,139.04);

\path[draw=drawColor,line width= 0.6pt,line join=round] ( 35.92, 35.65) --
	(139.04, 35.65);

\path[draw=drawColor,line width= 0.6pt,line join=round] ( 35.92, 59.78) --
	(139.04, 59.78);

\path[draw=drawColor,line width= 0.6pt,line join=round] ( 35.92, 83.92) --
	(139.04, 83.92);

\path[draw=drawColor,line width= 0.6pt,line join=round] ( 35.92,108.05) --
	(139.04,108.05);

\path[draw=drawColor,line width= 0.6pt,line join=round] ( 35.92,132.19) --
	(139.04,132.19);

\path[draw=drawColor,line width= 0.6pt,line join=round] ( 42.17, 30.72) --
	( 42.17,139.04);

\path[draw=drawColor,line width= 0.6pt,line join=round] ( 76.11, 30.72) --
	( 76.11,139.04);

\path[draw=drawColor,line width= 0.6pt,line join=round] (110.05, 30.72) --
	(110.05,139.04);
\definecolor{drawColor}{RGB}{0,0,255}
\definecolor{fillColor}{RGB}{0,0,255}

\path[draw=drawColor,line width= 0.6pt,line join=round,fill=fillColor] ( 40.60, 35.65) rectangle ( 43.73, 71.37);

\path[draw=drawColor,line width= 0.6pt,line join=round,fill=fillColor] ( 43.73, 35.65) rectangle ( 46.85,134.12);

\path[draw=drawColor,line width= 0.6pt,line join=round,fill=fillColor] ( 46.85, 35.65) rectangle ( 49.98,103.22);

\path[draw=drawColor,line width= 0.6pt,line join=round,fill=fillColor] ( 49.98, 35.65) rectangle ( 53.10, 81.02);

\path[draw=drawColor,line width= 0.6pt,line join=round,fill=fillColor] ( 53.10, 35.65) rectangle ( 56.23, 71.37);

\path[draw=drawColor,line width= 0.6pt,line join=round,fill=fillColor] ( 56.23, 35.65) rectangle ( 59.35, 62.68);

\path[draw=drawColor,line width= 0.6pt,line join=round,fill=fillColor] ( 59.35, 35.65) rectangle ( 62.48, 47.23);

\path[draw=drawColor,line width= 0.6pt,line join=round,fill=fillColor] ( 62.48, 35.65) rectangle ( 65.60, 46.27);

\path[draw=drawColor,line width= 0.6pt,line join=round,fill=fillColor] ( 65.60, 35.65) rectangle ( 68.73, 42.41);

\path[draw=drawColor,line width= 0.6pt,line join=round,fill=fillColor] ( 68.73, 35.65) rectangle ( 71.85, 43.37);

\path[draw=drawColor,line width= 0.6pt,line join=round,fill=fillColor] ( 71.85, 35.65) rectangle ( 74.98, 37.58);

\path[draw=drawColor,line width= 0.6pt,line join=round,fill=fillColor] ( 74.98, 35.65) rectangle ( 78.10, 38.54);

\path[draw=drawColor,line width= 0.6pt,line join=round,fill=fillColor] ( 78.10, 35.65) rectangle ( 81.23, 39.51);

\path[draw=drawColor,line width= 0.6pt,line join=round,fill=fillColor] ( 81.23, 35.65) rectangle ( 84.35, 39.51);

\path[draw=drawColor,line width= 0.6pt,line join=round,fill=fillColor] ( 84.35, 35.65) rectangle ( 87.48, 36.61);

\path[draw=drawColor,line width= 0.6pt,line join=round,fill=fillColor] ( 87.48, 35.65) rectangle ( 90.60, 39.51);

\path[draw=drawColor,line width= 0.6pt,line join=round,fill=fillColor] ( 90.60, 35.65) rectangle ( 93.73, 38.54);

\path[draw=drawColor,line width= 0.6pt,line join=round,fill=fillColor] ( 93.73, 35.65) rectangle ( 96.85, 35.65);

\path[draw=drawColor,line width= 0.6pt,line join=round,fill=fillColor] ( 96.85, 35.65) rectangle ( 99.98, 37.58);

\path[draw=drawColor,line width= 0.6pt,line join=round,fill=fillColor] ( 99.98, 35.65) rectangle (103.10, 38.54);

\path[draw=drawColor,line width= 0.6pt,line join=round,fill=fillColor] (103.10, 35.65) rectangle (106.23, 35.65);

\path[draw=drawColor,line width= 0.6pt,line join=round,fill=fillColor] (106.23, 35.65) rectangle (109.35, 35.65);

\path[draw=drawColor,line width= 0.6pt,line join=round,fill=fillColor] (109.35, 35.65) rectangle (112.48, 35.65);

\path[draw=drawColor,line width= 0.6pt,line join=round,fill=fillColor] (112.48, 35.65) rectangle (115.60, 35.65);

\path[draw=drawColor,line width= 0.6pt,line join=round,fill=fillColor] (115.60, 35.65) rectangle (118.73, 35.65);

\path[draw=drawColor,line width= 0.6pt,line join=round,fill=fillColor] (118.73, 35.65) rectangle (121.85, 35.65);

\path[draw=drawColor,line width= 0.6pt,line join=round,fill=fillColor] (121.85, 35.65) rectangle (124.98, 35.65);

\path[draw=drawColor,line width= 0.6pt,line join=round,fill=fillColor] (124.98, 35.65) rectangle (128.10, 35.65);

\path[draw=drawColor,line width= 0.6pt,line join=round,fill=fillColor] (128.10, 35.65) rectangle (131.23, 35.65);

\path[draw=drawColor,line width= 0.6pt,line join=round,fill=fillColor] (131.23, 35.65) rectangle (134.35, 36.61);
\definecolor{drawColor}{gray}{0.20}

\path[draw=drawColor,line width= 0.6pt,line join=round,line cap=round] ( 35.92, 30.72) rectangle (139.04,139.04);
\end{scope}
\begin{scope}
\path[clip] (  0.00,  0.00) rectangle (144.54,144.54);
\definecolor{drawColor}{gray}{0.30}

\node[text=drawColor,anchor=base east,inner sep=0pt, outer sep=0pt, scale=  0.88] at ( 30.97, 32.62) {0};

\node[text=drawColor,anchor=base east,inner sep=0pt, outer sep=0pt, scale=  0.88] at ( 30.97, 56.75) {25};

\node[text=drawColor,anchor=base east,inner sep=0pt, outer sep=0pt, scale=  0.88] at ( 30.97, 80.89) {50};

\node[text=drawColor,anchor=base east,inner sep=0pt, outer sep=0pt, scale=  0.88] at ( 30.97,105.02) {75};

\node[text=drawColor,anchor=base east,inner sep=0pt, outer sep=0pt, scale=  0.88] at ( 30.97,129.16) {100};
\end{scope}
\begin{scope}
\path[clip] (  0.00,  0.00) rectangle (144.54,144.54);
\definecolor{drawColor}{gray}{0.20}

\path[draw=drawColor,line width= 0.6pt,line join=round] ( 33.17, 35.65) --
	( 35.92, 35.65);

\path[draw=drawColor,line width= 0.6pt,line join=round] ( 33.17, 59.78) --
	( 35.92, 59.78);

\path[draw=drawColor,line width= 0.6pt,line join=round] ( 33.17, 83.92) --
	( 35.92, 83.92);

\path[draw=drawColor,line width= 0.6pt,line join=round] ( 33.17,108.05) --
	( 35.92,108.05);

\path[draw=drawColor,line width= 0.6pt,line join=round] ( 33.17,132.19) --
	( 35.92,132.19);
\end{scope}
\begin{scope}
\path[clip] (  0.00,  0.00) rectangle (144.54,144.54);
\definecolor{drawColor}{gray}{0.20}

\path[draw=drawColor,line width= 0.6pt,line join=round] ( 42.17, 27.97) --
	( 42.17, 30.72);

\path[draw=drawColor,line width= 0.6pt,line join=round] ( 76.11, 27.97) --
	( 76.11, 30.72);

\path[draw=drawColor,line width= 0.6pt,line join=round] (110.05, 27.97) --
	(110.05, 30.72);
\end{scope}
\begin{scope}
\path[clip] (  0.00,  0.00) rectangle (144.54,144.54);
\definecolor{drawColor}{gray}{0.30}

\node[text=drawColor,anchor=base,inner sep=0pt, outer sep=0pt, scale=  0.88] at ( 42.17, 19.71) {0};

\node[text=drawColor,anchor=base,inner sep=0pt, outer sep=0pt, scale=  0.88] at ( 76.11, 19.71) {100};

\node[text=drawColor,anchor=base,inner sep=0pt, outer sep=0pt, scale=  0.88] at (110.05, 19.71) {200};
\end{scope}
\begin{scope}
\path[clip] (  0.00,  0.00) rectangle (144.54,144.54);
\definecolor{drawColor}{RGB}{0,0,0}

\node[text=drawColor,anchor=base,inner sep=0pt, outer sep=0pt, scale=  1.10] at ( 87.48,  7.44) {Number of Accidentals};
\end{scope}
\begin{scope}
\path[clip] (  0.00,  0.00) rectangle (144.54,144.54);
\definecolor{drawColor}{RGB}{0,0,0}

\node[text=drawColor,rotate= 90.00,anchor=base,inner sep=0pt, outer sep=0pt, scale=  1.10] at ( 13.08, 84.88) {Number of Shoes};
\end{scope}
\end{tikzpicture}\label{accidentalcounts}} \hspace{1cm}
\subfloat[]{\includegraphics[height = 2in]{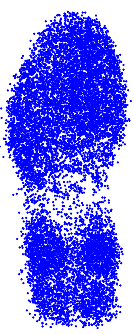}\label{accidentallocations}}
\caption{(a) is a histogram summarizing the number of accidentals on each shoe in JESA. (b) illustrates the locations of these accidentals with points.}
\end{figure}
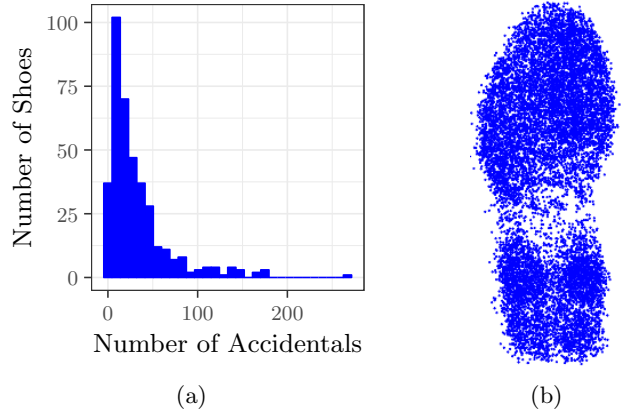

The number of accidentals, and their locations, varies from shoe to shoe. Figure~\ref{accidentalcounts} provides a histogram of the number of accidentals on each shoe. The distribution is heavily skewed to the right-- the median number of accidentals is 20, whereas the mean is 33, and the maximum is 268.

Figure~\ref{accidentallocations} aggregates the coordinates of all accidentals recorded in the JESA database. Its similarity to that of Figure~\ref{aggregatedcontacts} is consistent with the intuition that accidentals should appear more frequently in areas of the shoe which are part of the contact surface. However, not all accidental locations fall on the sole with contact surface. Of the accidentals in JESA, about 12 percent of them occur in grid points without contact. Therefore, a robust model should be able to assign probability to situations in which accidentals do not occur directly on the contact surface. Examples of shoes in JESA for which accidentals occur away from the contact surface are available in the Appendix (Figure~\ref{offcontacts}). 
Following \cite{damary2018dependence}, we exclude rift-type accidentals from our analysis because they occur only on specific type of shoe tread, making their spatial distribution markedly different than the more frequently occurring types of accidentals (e.g. hole or scratch).

\subsection{Existing Models for the Distribution of Accidentals}\label{problem_form}

Going forward, we use the shorthand $x^s | \mathcal{C}^s$ to refer the distribution of accidental locations $x^s$ on a shoe $s$ with contact surface $\mathcal{C}^s$, with $\mathcal{C}^s = \mathcal{C}^y$ referring to the distribution required to compute the RMP (\ref{rmpv}). For easy comparison of existing models in the literature with the approach we develop, we use a unified notation.

We begin by treating each $x^s | \mathcal{C}^s$ as a draw from a 2-dimensional spatial point process \citep{daley2007introduction} over the standardized space $[0,100] \times [0,200]$. For our model, we make three additional assumptions regarding the structure of these point processes: (1) the individual accidentals  $(x^s_n)_{n=1\ldots N_s}$ are exchangeable, (2)  the marginal distribution of each $x^s_n$ is independent of the total number of accidentals $N_s$, and (3) the distribution of $x^s$ depends on $s$ only through the contact surface $\mathcal{C}^s$. The first two assumptions are common in the literature, whereas the third is unique to our model because we are first to incorporate the contact surface. 

Following assumptions (1) and (2), $(x^s_n)_{n=1\ldots N_s}$ can be treated as independent draws from a random probability measure $\Lambda_s$ on $[0,100] \times [0,200]$. The literature has mostly focused on universal models for $\Lambda_s$, assuming a fixed $\Lambda$ that is common to all shoes $s \in \pop$. \cite{stone2006footwear} proposed a uniform model for $\Lambda$, i.e. $\Lambda \propto 1$. This assumption has been criticized for its lack of empirical support, as noted by \citet{president2016report}. \cite{yekutieli2012expert} instead inferred $\Lambda$ using a kernel density estimator on the accidentals in JESA (Section~\ref{data}). \cite{speir2016quantifying} applied a similar histogram estimator to a different database, yielding comparable results. 

Because estimating a single $\Lambda$ does not allow for conditioning on class characteristics or wear, these approaches implicitly assume that a shoe's accidental locations are independent of its contact surface. Evidence against this assumption was provided by \cite{damary2018dependence}; their analysis of multiple replicates of three different tread patterns appearing in the JESA database revealed that different tread patterns tend to yield different accidental distributions. Therefore, having distinct $\Lambda_s$ that depend on $\mathcal{C}^s$ seems more appropriate, serving as the motivation for our assumption (3) above.

We encode assumption (3) in our model by explicitly treating each $\Lambda_s$ as a draw from a distribution $G_{\mathcal{C}^s}$. As the notation suggests, $G_{\mathcal{C}^s} = G_{\mathcal{C}^{s'}}$ if $\mathcal{C}^s = \mathcal{C}^{s'}$, but the distributions of $\Lambda_s$ and $\Lambda_{s'}$ can differ otherwise. Other works have followed a similar line of thought by restricting analysis to a single type of shoe at a time \citep{adair2007mount, petraco2010statistical, wilson2012comparison}. In each of those studies, several replicates of the exact same pair of shoes were worn independently for a period of time, after which their accidental locations were annotated, analyzed, and compared. This allowed for the identification of common trends for one specific type of shoe. Though such data is ideal for modeling $G_{\mathcal{C}^s}$, the approach cannot be practically scaled to all types of shoes. Collecting multiple annotated observations for all given tread patterns is prohibitively expensive. In addition, the project would have to continue in perpetuity, continually updating the database to account for the ever-growing list of footwear styles and brands.

For this reason, we propose a more general and scalable approach in our modeling of $\Lambda_s$. Instead of developing independent models $G_{\mathcal{C}^s}$ for each unique contact surface, we propose a Bayesian hierarchical model to pool information across many contact surfaces at once. Let $\mathbb{C}$ denote the space of possible contact surfaces. Our goal is to infer the entire family of distributions $G = (G_{\mathcal{C}})_{\mathcal{C} \in \mathbb{C}}$ as a single model, treating a shoe's contact surface $\mathcal{C}$ as a covariate. This joint modeling approach helps to leverage the information available in heterogeneous databases --- in our case the JESA database --- to identify the relationship between the contact surface and accidental locations and to capture commonalities that span across many shoe types.

Let $\jesa$ denote a set of available shoes (e.g. JESA) used to infer $G$. Then $(\Lambda_s)_{s \in \jesa}$ is a vector of dependent random probability measures, with the dependence between them induced by a hierarchical model on $G$. We now review existing approaches for modeling vectors of dependent probability measures, limiting our discussion to that which is most relevant to our model. We defer discussion of additional related work to the Appendix (\S\ref{relatedwork2}).

\subsection{Random Vectors of Dependent Probability Measures}\label{sec:crm}

Over the years, there has been a broad interest in modeling dependent probability measures, especially via nonparametric Bayes \citep{hjort2010bayesian, foti2015survey}. The approach we use to model $(\Lambda_s)_{s \in \jesa}$ in this paper is not fully nonparametric, but it is a finite-resolution approximation of one. Thus, it is natural to frame our review within the nonparametric Bayesian literature. 

The canonical Bayesian nonparametric approach to modeling a measure $\mu$ on a space $\Omega$ is to treat it as a random draw from some subclass of measures on $\Omega$. Completely random measures \citep{kingman1967completely} are an especially tractable subclass of random measures that are composed of a (possibly countably infinite) collection of weighted atoms in $\Omega$. We use $(\theta_i)_{i=1,...,\infty} \in \Omega^{\infty}$ to denote the locations of the atoms of the completely random measure $\mu$, and $(w_i)_{i=1, \ldots, \infty} \in \mathbb{R}_+^{\infty}$ to denote the corresponding (non-negative) atom weights. The defining feature of a completely random measure is that, for any disjoint subsets $\Omega_1, \Omega_2  \subset \Omega$, $\mu(\Omega_1)$ is independent of $\mu(\Omega_2)$ (complete randomness). An accessible review of completely random measures as they pertain to statistical modeling is available in \cite{jordan2010hierarchical}. 

For our purposes, we are interested in atomic measures that do not necessarily satisfy the complete randomness assumption. In particular, we are interested in atomic random \emph{probability} measures --- random measures $\mu$ consisting of atoms such that $\mu(\Omega) = 1$. Any finite atomic random measure can be converted to a probability measure via normalization. For instance, a normalized completely random measure takes the form
\begin{align}\label{normalized}
\bar{\mu}(\cdot) &= \frac{\sum_{i=1}^{\infty} w_i \delta_{\theta_i}(\cdot)}{ \sum_{i=1}^{\infty} w_i }.
\end{align}
where $w_i, \theta_i$ are defined analogously to above. The strength of atomic probability measures is that they can be convolved with probability kernels to define mixture models for densities (e.g. \cite{escobar1995bayesian}, \cite{rasmussen2000infinite}). Each atom acts as its own mixture component, providing a framework that is flexible and computationally tractable.

Rather than a single normalized random measure, we are concerned with a vector of dependent random probability measures $(\Lambda_s)_{s \in \jesa}$ that can capture commonalities across all shoes in JESA. Particularly relevant to our work is the recently proposed normalized compound random measure framework (NCoRM) of \cite{griffin2017compound}, which formulates the vector of random probability measures $\mu_1, \ldots, \mu_K$ on $\Omega$ as
\begin{align}
\mu_{k}(\cdot) &= \frac{\sum_{i=1}^{\infty} m^k_{i} w_{i} \delta_{\theta_i}(\cdot)}{\sum_{i=1}^{\infty} m^k_{i} w_{i}}
\end{align}
where $(\theta_i, w_i)_{i=1,...,\infty}$ are drawn as in a single completely random measure and $(m^k_{i})_{i=1,\ldots, \infty}$ are iid random ``score'' variables for $k=1,\ldots, K$, following a distribution $\rho$, that up-weight or down-weight the shared set of atoms defined by the $(\theta_i, w_i)$ for each of the $\mu_k$'s. The distribution of the scores controls the strength of the dependence, with much of the exposition in \cite{griffin2017compound} devoted to gamma distributions due to their computational tractability. We use the idea of scoring in normalized atomic random measures to develop our model. However, modifications must be made.

The NCoRM approach as described in \cite{griffin2017compound} was developed for exchangeable vectors of random probability measures. However, exchangeability does not hold when each measure has an associated covariate (as we have in the contact surfaces $\mathcal{C}^s$). For this reason, we generalize the idea of ``scoring'' from NCoRMs to the non-exchangeable setting, allowing us to incorporate covariate information. It is worth noting that \cite{griffin2018modelling} also generalizes the NCoRM framework to a non-exchangeable regression framework, but differently than we do here.


\section{Model}\label{model}

Recall that for a given shoe $s \in \jesa$, we have assumed each accidental location $x_n^s$ is drawn independently from a probability measure $\Lambda_s$ on $[0,100] \times [0,200]$ where $\Lambda_s$ itself is randomly drawn from a distribution $G_{\mathcal{C}^s}$ that depends on the contact surface $\mathcal{C}^s \in \mathbb{C}$. Because it is impractical to independently model $G_{\mathcal{C}}$ for all possible $\mathcal{C} \in \mathbb{C}$, we develop a hierarchical model to jointly infer all entries of $G$, treating each $\mathcal{C} \in \mathbb{C}$ as a high-dimensional spatial covariate. 

Before specifying how we model the family of distributions $G$, it is useful to first address the limited precision of the data. As per \S\ref{data}, the contact surface variables $\mathcal{C} \in \mathbb{C}$ are defined on a discrete 200 by 100 equally-spaced grid over $[0,100] \times [0,200]$. We use $\mathcal{A}$ to denote the set of entries in this grid:
\begin{align}
\mathcal{A} = \left\{(a_1,a_2): a_1 \in \{1, \ldots, 100\}, a_2 \in \{1, \ldots, 200\}\right\}
\end{align} 
with gridpoint $(a_1, a_2) \in \mathcal{A}$ corresponding to the area $(a_1 - 1, a_1] \times (a_2 - 1, a_2]$ in $[0,100] \times [0,200]$. We restrict our model for $\Lambda_s$ to have the same resolution as $\mathcal{A}$ by discretizing $\Lambda_s$ to be a piece-wise constant over each gridpoint in $\mathcal{A}$. This reduced resolution provides computational advantages, simplifies interpretation, and guards against overfitting. Further discussion of the discretization is available in  the Appendix (\S\ref{discretization}).

After discretization, each $\Lambda_s$ can be characterized by the values it takes at the grid points in $\mathcal{A}$, and each $G_{\mathcal{C}} \in G$ can be characterized by the multivariate distribution it assigns to those grid points. This provides a natural representation for parametrizing our model --- we view $G$ as a family of distributions over the 20000-dimensional simplex indexed by $\mathbb{C}$, with each $G_{\mathcal{C}^s}$ characterized by the joint distribution it defines over the vector of values in the probability measure $\Lambda_s|\mathcal{C}^s$. It is most straightforward to describe $G$ in terms of the generative process it assigns to a generic $\Lambda_s|\mathcal{C}^s$, as we do below.

\subsection{Parameterization of $\Lambda_s$}

We model each measure $\Lambda_s \sim G_{\mathcal{C}^s}$ as the convolution of a normalized random atomic measure $\mu_s$ with a two dimensional piece-wise constant probability kernel $\kernel$. We define $\mu_s$ to consist of 20000 atoms at fixed locations --- one for each gridpoint in $\mathcal{A}$. To model the weights of each of these atoms, we generalize the NCoRM scoring technique of \cite{griffin2017compound} to incorporate the covariate information in $\mathcal{C}^s$, and to allow for spatial dependence between atom weights.

For each $a \in \mathcal{A}$, we define the distribution of $\mu_s | \mathcal{C}^s$ as
\begin{align}
\mu_s(a) &= \frac{\common_{a} m_{a}^{s} }{\sum_{b \in \mathcal{A}} \common_{b} m_{b}^{s} }
                                             = \frac{\common_{a} \noise_{a}^{s} \contact^s_{a} }{\sum_{b \in \mathcal{A}} \common_{b} \noise_{b}^{s} \contact_{b} }\label{gofurther}.
\end{align}
Here, $(\common_{a})_{a \in \mathcal{A}}$ are parameters common to all $G$, and $(m_{a}^s)_{a \in \mathcal{A}}$ are random shoe-specific location-specific scores applied to the weights of the atoms. The scores further decompose into two components: $m_{a}^{s} = \noise_{a}^s \contact_{a}^s$, with $\noise_{a}$ representing ``traditional'' scores as in NCoRM (assumed to be independent for all shoes and all locations), and $\contact^s_{a}$ representing contact-dependent scores --- variables that depend on the nearby configuration of $\mathcal{C}^s$. We model the traditional scores as independent draws from $\rho_q =$ Gamma($q$, 1). The contact-dependent scores $\contact^s_{b}$ are treated as parameters, defined as follows. 

\begin{figure}[!h]
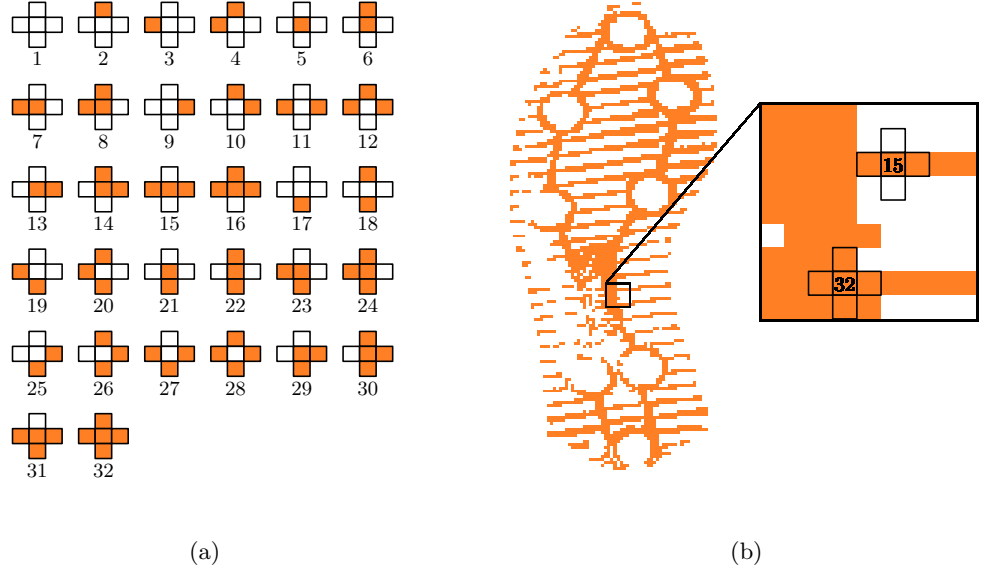
 
\centering
  \subfloat[]{\input{phi_options2.tex} \label{phia}} 
    \subfloat[]{\input{phi_zoom.tex} \label{phib}}\\ 
\caption{(a) provides a list of the possible shapes the contact surface can take around an atom, accompanied by the index in $\contact \in [0,1]^{32}$ to which it corresponds. (b) zooms in on an example shoe's contact surface (zoomed region outlined in black) to demonstrate the $\contact^s_a$ value of two example locations.} \label{phi} 
\end{figure}

Let $\contact \in [0,1]^{32}$. For all $a \in \mathcal{A}$, $s \in \jesa$ define
\begin{align}
\contact^s_{a} &= \contact_{r^s_{a}} \text{ where}\\
r^s_{a} &= 1+ \sum_{i=-1}^{1} \sum_{j= -1}^1 2^{3 + i + 2j} \mathcal{C}^s_{a + (i,j)} I(||(i,j)||^2 \leq 1) \label{rsa}.
\end{align}
By this formulation, $\contact^s_{a}$ takes one of $2^5 = 32$ values depending on the value of the contact surface at the gridpoints surrounding $a$. For instance, if $a$ is completely surrounded by contact surface, i.e.
\begin{align}
\mathcal{C}_{a + (-1,0)} = \mathcal{C}_{a + (0,-1)} = \mathcal{C}_{a} = \mathcal{C}_{a + (1,0)} = \mathcal{C}_{a + (0,1)} = 1,
\end{align} 
then $\contact^s_{a} = \contact_{32}$. Similarly, if $a$ is in an area devoid of contact surface, i.e.
\begin{align}
\mathcal{C}_{a + (-1,0)} = \mathcal{C}_{a + (0,-1)} = \mathcal{C}_{a} = \mathcal{C}_{a + (1,0)} = \mathcal{C}_{a + (0,1)} = 0,
\end{align} 
then $\contact^s_{a} = \contact_{1}$. A demonstration of the possible configurations is provided in Figure~\ref{phia} along with an depiction of $r^s_{a}$ for two $a \in \mathcal{A}$ in Figure~\ref{phib}.

Before specifying the functional form for the kernel $\kernel$ (which smooths the atom weights), let us first interpret of the various components that define the atoms weights for $\mu_s$ in the context of the shoe sole and accidentals. \\
The weights are the normalized product of three components: 
\begin{enumerate}
\item $\contact$, which specifies the impact of a gridpoint's surrounding contact surface on the relative likelihood of accidental occurrence, 
\item $\common$, which specifies the impact of the position of a gridpoint's spatial coordinates on the relative likelihood of accidental occurrence, and
\item $\rho_q$ (parameterized by $q$), which specifies the variability across shoes of each gridpoint's relative probability of accidental occurrence, controlling for position and contact surface.
\end{enumerate}
Essentially, the parameters $\contact$ and $\common$ control the mean of $\mu_s$, whereas its variance depends on the $\noise_{a}^s$ scores --- distributed according to $\rho_q$. These choices are in-line with a common belief in forensic footwear analysis --- that the locations of accidentals tend to follow a spatially inhomogeneous distribution across the shoe sole (captured by $\common$), and that some areas are more likely to be affected than others depending on their contact with the ground (captured by $\contact$). We model each of $\contact$, $\common$, and $q$ as global parameters, assuming they take the same value for all shoes JESA. 

The random shoe-specific errors $\noise^s$ capture deviations from this common trend. The coefficient of variation of $\rho_q$ --- given by $q^{-1/2}$ --- indicates the strength of the deviations. The smaller the value of $q$, the larger the variation of $\mu_s$ around its mean.

\begin{figure}[!h]
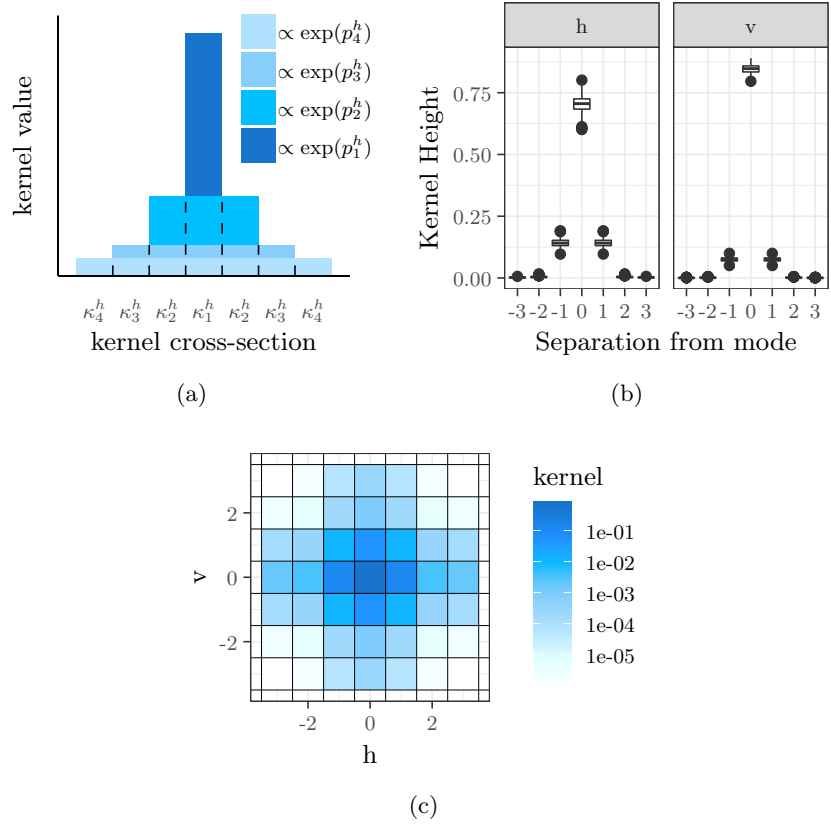

  \centering 
  \subfloat[]{ 
\begin{tikzpicture}[x=1pt,y=1pt]
\definecolor{fillColor}{RGB}{255,255,255}
\path[use as bounding box,fill=fillColor,fill opacity=0.00] (0,0) rectangle (144.54,144.54);
\begin{scope}
\path[clip] ( 17.77, 30.72) rectangle (139.04,139.04);
\definecolor{fillColor}{RGB}{176,226,255}

\path[fill=fillColor] ( 30.17, 35.65) rectangle (126.64, 42.42);
\definecolor{fillColor}{RGB}{135,206,250}

\path[fill=fillColor] ( 43.95, 42.42) rectangle (112.86, 47.34);
\definecolor{fillColor}{RGB}{0,191,255}

\path[fill=fillColor] ( 57.73, 47.34) rectangle ( 99.08, 65.80);
\definecolor{fillColor}{RGB}{24,116,205}

\path[fill=fillColor] ( 71.51, 65.80) rectangle ( 85.30,127.35);
\definecolor{drawColor}{RGB}{0,0,0}

\path[draw=drawColor,line width= 0.6pt,dash pattern=on 4pt off 4pt ,line join=round] (126.64, 35.65) -- (126.64, 35.65);

\path[draw=drawColor,line width= 0.6pt,dash pattern=on 4pt off 4pt ,line join=round] (112.86, 35.65) -- (112.86, 42.42);

\path[draw=drawColor,line width= 0.6pt,dash pattern=on 4pt off 4pt ,line join=round] ( 99.08, 35.65) -- ( 99.08, 47.34);

\path[draw=drawColor,line width= 0.6pt,dash pattern=on 4pt off 4pt ,line join=round] ( 85.30, 35.65) -- ( 85.30, 65.80);

\path[draw=drawColor,line width= 0.6pt,dash pattern=on 4pt off 4pt ,line join=round] ( 30.17, 35.65) -- ( 30.17, 35.65);

\path[draw=drawColor,line width= 0.6pt,dash pattern=on 4pt off 4pt ,line join=round] ( 43.95, 35.65) -- ( 43.95, 42.42);

\path[draw=drawColor,line width= 0.6pt,dash pattern=on 4pt off 4pt ,line join=round] ( 57.73, 35.65) -- ( 57.73, 47.34);

\path[draw=drawColor,line width= 0.6pt,dash pattern=on 4pt off 4pt ,line join=round] ( 71.51, 35.65) -- ( 71.51, 65.80);

\path[draw=drawColor,line width= 0.6pt,line join=round] ( 23.28, 35.65) -- (133.53, 35.65);

\path[draw=drawColor,line width= 0.6pt,line join=round] ( 23.28, 35.65) -- (133.53, 35.65);

\path[draw=drawColor,line width= 0.6pt,line join=round] ( 23.28, 35.65) -- (133.53, 35.65);

\path[draw=drawColor,line width= 0.6pt,line join=round] ( 23.28, 35.65) -- (133.53, 35.65);

\path[draw=drawColor,line width= 0.6pt,line join=round] ( 23.28, 35.65) -- ( 23.28,134.12);

\path[draw=drawColor,line width= 0.6pt,line join=round] ( 23.28, 35.65) -- ( 23.28,134.12);

\path[draw=drawColor,line width= 0.6pt,line join=round] ( 23.28, 35.65) -- ( 23.28,134.12);

\path[draw=drawColor,line width= 0.6pt,line join=round] ( 23.28, 35.65) -- ( 23.28,134.12);
\end{scope}
\begin{scope}
\path[clip] (  0.00,  0.00) rectangle (144.54,144.54);
\definecolor{drawColor}{gray}{0.30}

\node[text=drawColor,anchor=base,inner sep=0pt, outer sep=0pt, scale=  0.88] at ( 37.06, 19.71) {$\kappa_4^h$};

\node[text=drawColor,anchor=base,inner sep=0pt, outer sep=0pt, scale=  0.88] at ( 50.84, 19.71) {$\kappa_3^h$};

\node[text=drawColor,anchor=base,inner sep=0pt, outer sep=0pt, scale=  0.88] at ( 64.62, 19.71) {$\kappa_2^h$};

\node[text=drawColor,anchor=base,inner sep=0pt, outer sep=0pt, scale=  0.88] at ( 78.40, 19.71) {$\kappa_1^h$};

\node[text=drawColor,anchor=base,inner sep=0pt, outer sep=0pt, scale=  0.88] at ( 92.19, 19.71) {$\kappa_2^h$};

\node[text=drawColor,anchor=base,inner sep=0pt, outer sep=0pt, scale=  0.88] at (105.97, 19.71) {$\kappa_3^h$};

\node[text=drawColor,anchor=base,inner sep=0pt, outer sep=0pt, scale=  0.88] at (119.75, 19.71) {$\kappa_4^h$};
\end{scope}
\begin{scope}
\path[clip] (  0.00,  0.00) rectangle (144.54,144.54);
\definecolor{drawColor}{RGB}{0,0,0}

\node[text=drawColor,anchor=base,inner sep=0pt, outer sep=0pt, scale=  1.10] at ( 78.40,  7.44) {kernel cross-section};
\end{scope}
\begin{scope}
\path[clip] (  0.00,  0.00) rectangle (144.54,144.54);
\definecolor{drawColor}{RGB}{0,0,0}

\node[text=drawColor,rotate= 90.00,anchor=base,inner sep=0pt, outer sep=0pt, scale=  1.10] at ( 13.08, 84.88) {kernel value};
\end{scope}
\begin{scope}
\path[clip] (  0.00,  0.00) rectangle (144.54,144.54);
\definecolor{fillColor}{RGB}{255,255,255}

\path[fill=fillColor] ( 86.41, 72.14) rectangle (150.43,140.95);
\end{scope}
\begin{scope}
\path[clip] (  0.00,  0.00) rectangle (144.54,144.54);
\definecolor{fillColor}{RGB}{255,255,255}

\path[fill=fillColor] ( 91.91,121.00) rectangle (106.37,135.45);
\end{scope}
\begin{scope}
\path[clip] (  0.00,  0.00) rectangle (144.54,144.54);
\definecolor{fillColor}{RGB}{176,226,255}

\path[fill=fillColor] ( 92.63,121.71) rectangle (105.66,134.74);
\end{scope}
\begin{scope}
\path[clip] (  0.00,  0.00) rectangle (144.54,144.54);
\definecolor{fillColor}{RGB}{255,255,255}

\path[fill=fillColor] ( 91.91,106.55) rectangle (106.37,121.00);
\end{scope}
\begin{scope}
\path[clip] (  0.00,  0.00) rectangle (144.54,144.54);
\definecolor{fillColor}{RGB}{135,206,250}

\path[fill=fillColor] ( 92.63,107.26) rectangle (105.66,120.29);
\end{scope}
\begin{scope}
\path[clip] (  0.00,  0.00) rectangle (144.54,144.54);
\definecolor{fillColor}{RGB}{255,255,255}

\path[fill=fillColor] ( 91.91, 92.09) rectangle (106.37,106.55);
\end{scope}
\begin{scope}
\path[clip] (  0.00,  0.00) rectangle (144.54,144.54);
\definecolor{fillColor}{RGB}{0,191,255}

\path[fill=fillColor] ( 92.63, 92.80) rectangle (105.66,105.83);
\end{scope}
\begin{scope}
\path[clip] (  0.00,  0.00) rectangle (144.54,144.54);
\definecolor{fillColor}{RGB}{255,255,255}

\path[fill=fillColor] ( 91.91, 77.64) rectangle (106.37, 92.09);
\end{scope}
\begin{scope}
\path[clip] (  0.00,  0.00) rectangle (144.54,144.54);
\definecolor{fillColor}{RGB}{24,116,205}

\path[fill=fillColor] ( 92.63, 78.35) rectangle (105.66, 91.38);
\end{scope}
\begin{scope}
\path[clip] (  0.00,  0.00) rectangle (144.54,144.54);
\definecolor{drawColor}{RGB}{0,0,0}

\node[text=drawColor,anchor=base west,inner sep=0pt, outer sep=0pt, scale=  0.88] at (106.37,125.20) {$\propto \exp(p_4^h)$};
\end{scope}
\begin{scope}
\path[clip] (  0.00,  0.00) rectangle (144.54,144.54);
\definecolor{drawColor}{RGB}{0,0,0}

\node[text=drawColor,anchor=base west,inner sep=0pt, outer sep=0pt, scale=  0.88] at (106.37,110.74) {$\propto \exp(p_3^h)$};
\end{scope}
\begin{scope}
\path[clip] (  0.00,  0.00) rectangle (144.54,144.54);
\definecolor{drawColor}{RGB}{0,0,0}

\node[text=drawColor,anchor=base west,inner sep=0pt, outer sep=0pt, scale=  0.88] at (106.37, 96.29) {$\propto \exp(p_2^h)$};
\end{scope}
\begin{scope}
\path[clip] (  0.00,  0.00) rectangle (144.54,144.54);
\definecolor{drawColor}{RGB}{0,0,0}

\node[text=drawColor,anchor=base west,inner sep=0pt, outer sep=0pt, scale=  0.88] at (106.37, 81.83) {$\propto \exp(p_1^h)$};
\end{scope}
\end{tikzpicture} \label{kernela}}
      \subfloat[]{ \input{kernel_posterior.tex} \label{kernelb}} \\
       \subfloat[]{ \input{kernel2dposterior.tex} \label{kernelc}}
\caption{(a) illustrates the tiered cake parametrization of $\kappa^h$. Each uniquely colored tier is proportional to the corresponding $\exp{p^h_i}$, with the dotted lines depicting how the cake is sliced that form each $\kappa^h_i$.
(b) demonstrates the posterior fit of $\kappa^h$ and $\kappa^v$ using symmetrically arranged boxplots.
(c) depicts the posterior mean of $\kernel(i,j) = \kappa^h_i \kappa^v_j$ centered $(0,0)$. The decay in the $h$ direction controlled by $\kappa^h$, and the decay in the vertical direction ($v$) controlled by $\kappa^v$, with the hue changing according to a logarithmic scale.} \label{kernel}
\end{figure}

Finally, we convolve all atoms in all $\mu_s$ with a kernel $\kernel$ to obtain $\Lambda_s$. The kernel is parameterized to smooth the weights across nearby atoms. Recognizing that the smoothing should be local (Appendix \S\ref{discretization}), we define the kernel $\kernel$ to have finite support, symmetrically redistributing the mass over a window extending three grid points from $a$ in all four axis-aligned directions. Figure~\ref{kernela} illustrates the shape of the probability kernel. We refer to this parameterization as the \emph{tiered cake} representation due to the resultant kernel resembling a tiered cake with $p^\alpha$ controlling the size of each tier.

We parameterize $\kernel$ as a function $\kernel: \{-3,\ldots, 3\}^2 \rightarrow [0,1]$ such that
\begin{align}
\kernel(i,j) &= \kappa^h_{1+|i|} \kappa_{1+|j|}^v. \label{kernelequation}
\end{align}
Here, $\kappa^h, \kappa^v \in [0,1]^4$ define independent symmetric kernels in the horizontal and vertical directions, and $\kernel$ is their composition. To ensure that $\kappa^{h}$ and $\kappa^v$ are unimodal probability kernels, we further re-parameterize them as
\begin{align}
\kappa^{\alpha}_i = \frac{\sum_{j=i}^4 \exp(p^\alpha_j)/(2j-1)}{\sum_{j=1}^4\exp(p^\alpha_j) },
\end{align}
for $i =1,\ldots,4$, $\alpha = v,h$, and each $p^{\alpha} \in \mathbb{R}^4$. Note that our fitted results (Figure~\ref{kernelb}) indicate that extending the window for three grid points appears to be excessive, but parameterizing three allowed for such a discovery.
Going forward, we will often suppress the $p^{\alpha}$ parameterization to make the presentation more concise, instead relying on the $\kappa^{\alpha}$ representation.

\subsection{Model Summary and Prior}

Having parametrized $G$, we now formulate the full hierarchical Bayesian model. Let $\Theta$ denote the concatenation of the global parameters $\contact$, $\common$, $q$,  $p^h$, and $p^v$. Our prior distribution on $\Theta$ is the composition of independent priors on its entries. Letting $\text{MVN}(0, 4 I_4)$ denote 4-dimensional isotropic Gaussian distribution with variance 4, and $\text{MVLN}(0, \Sigma)$ denote a multivariate log normal distribution with mean parameter 0 and precision matrix $\Sigma$, the following provides a bird's eye view of the model via the generative process of the JESA data given $(\mathcal{C}_s, N_s)_{s \in \jesa}$:\\

\noindent \textbf{Step 1:} Generate global parameters:
\begin{align*}
  q &\sim \text{Gamma}(2,2), &\common_E &\sim \text{MVLN}( 0, \Sigma),\\
 \contact & \sim \text{unif}([0,1]^{32}), & p^h, p^v & \sim \text{MVN}(0, 4 I_4).
  \end{align*}
 \textbf{Step 2:} Generate the densities $\Lambda_s \sim G_{\mathcal{C}^s}$ for $s \in \jesa$:
 \begin{align*}
\text{For } a \in \mathcal{A}:\\
 &  \noise^j_{a} \sim \text{Gamma}(q,1), \\
&\Lambda_s(a) = \sum_{-3 \leq i,j \leq 3} \kappa^h_{1 + |i|} \kappa^v_{1+ |j|} \frac{\common_{a + (i,j)} \noise_{a + (i,j)}^{s} \contact^s_{a + (i,j)} }{\sum_{a' \in \mathcal{A}'} \common_{a'} \noise_{a'}^{s} \contact^s_{a'} }.
\end{align*}
 \textbf{Step 3:} Generate the accidental Locations $x^s$ for $s \in \jesa$.
 \begin{align*}
 \text{For } n =1,\ldots, N_s: \text{ generate } x^s_n \sim \Lambda_s.
\end{align*}

Our prior on $\common$ in Step~1 uses a coarsened representation, the details of which are provided in the Appendix (\S\ref{commondetails} and Figure~\ref{Jgrid}). The vector $\common_E$ denotes the subvector of unique values after the coarsening. The scales for all of the priors were chosen based on the range of possible behaviors we expected in the model. For example, the variances of 4 for $p^h$ and $p^v$ were chosen to strike a balance: too small of a variance concentrates the kernel around its geometric decaying mean, and too large of a variance places most of the prior density kernels that are essentially step functions. For $\contact \in [0,1]^{32}$, the upper bound on the uniforms is arbitrary --- the likelihood in (\ref{marginal1}) is invariant to scalings of $\contact$ due to the normalization of $\mu_s$. The rate of $\rho_q$ is fixed at 1 for the same reason.

\section{Computation}\label{computation}

There are two key computational challenges associated with our model.
\begin{enumerate}
\item How do we efficiently compute the posterior of $\Theta$? \label{task1}
\item How do we efficiently compute the density of an observed set of accidentals $x^s$ given $\mathcal{C}^s$? \label{task2}
\end{enumerate}
Task 1 (addressed in \S\ref{MCMC}) arises when fitting our model to the JESA data, and task 2 (addressed in \S\ref{IS}) arises when evaluating models. Before describing our strategies for addressing these tasks, we develop a trick to compute the likelihood of $x^s\in ([0,100] \times [0,200])^{N_s}$ given $\mathcal{C}^s$ for a given $\Theta$. 

The raw likelihood takes the form
\begin{align}
p(x^s| \mathcal{C}^s; \Theta) &=  \int \prod_{n=1}^{N_s} \Lambda(x^s_n) G_{\mathcal{C}^s}(\text{d} \Lambda)\\
&= \int \prod_{n=1}^{N_s} \sum_{-3 \leq i,j \leq 3} \kappa^h_{1 + |i|} \kappa^v_{1+ |j|} \frac{\common_{x^s_n +(i,j)} \noise_{x^s_n + (i,j)}^{s} \contact^s_{x^s_n + (i,j)} }{\sum_{a \in \mathcal{A}} \common_{a} \noise_{a}^{s} \contact^s_{a} } \text{d}\rho(\noise^s).  \label{marginal1}
\end{align}
In a slight abuse of notation, we have overloaded $x^s_n$ to also denote the atom $a \in \mathcal{A}$ to which the real-valued $x^s_n \in [0,100]\times [0,200]$ is associated. At first glance, the $|\mathcal{A}|$-dimensional integral over the $\noise^s$ variable in (\ref{marginal1}) appears to be computationally intractable. It has no closed form, and is too high dimensional to efficiently compute using quadrature or generic Monte Carlo algorithms. To overcome this problem, we introduce auxiliary variables.

For each accidental location $x^s_{n}$ on shoe $s \in \jesa$, we define $Z^s_{n}$ by
\begin{align}
\mathbb{P}(Z^s_n = x^s_n + (i,j) \mid \kappa^h, \kappa^v) = \kernel(i,j) = \kappa^h_{1 + |i|} \kappa^v_{1 + |j|} \label{distz},
\end{align}
with $\kappa^v, \kappa^h$ being the kernel parameters as defined in (\ref{kernelequation}), and each $x^s_n \in \mathcal{A}$. We use the shorthand $Z^s$ to refer to the collection $(Z^s_n)_{1 \leq n \leq N_s}$ and use $C^s_a$ to denote the number of times each $a \in \mathcal{A}$ occurs in $Z^s$. We also introduce the auxiliary variables
\begin{align}
u^s \sim \text{Gamma}\left( N_s, \sum_{a \in \mathcal{A}} \common_{a} \noise_{a}^{s} \contact^s_{a}\right),
\end{align}
with $\text{Gamma}(\alpha, \beta)$ denoting a gamma distribution with shape $\alpha$ and rate $\beta$. We can now analytically marginalize the $\noise^s$ variables to obtain
\begin{align}\label{marginal2}
p(x^s| \mathcal{C}^s; \Theta)  &=   \int_{0}^{\infty} \frac{u^{N_s - 1}}{\Gamma(N_s)}\mathbb{E} \left(  \frac{1}{\Gamma(q)^{|\mathcal{A}|}}\prod_{a \in \mathcal{A}} \frac{ \Gamma(q + C_a^s) \left(\common_{a} \contact^s_{a}\right)^{C_a}}{\left(u^s \common_{a}\contact^s_{a} + 1 \right) ^{q + C^s_a}} \right) \text{d}u^s,
\end{align}
where $\Gamma(\cdot)$ denotes the gamma function and $\mathbb{E}$ denotes an expectation taken with respect to the distribution of $Z^s$ as given in (\ref{distz}). By swapping (\ref{marginal1}) for (\ref{marginal2}), we have exchanged a $|\mathcal{A}|$-dimensional integral over $\noise^s$ for a more tractable one dimensional integral. The full derivation of moving from (\ref{marginal1}) to (\ref{marginal2}) is provided in the Appendix (\S\ref{marginalization}).

This new expression for the marginal likelihood (\ref{marginal2}) enables us to address challenges (1) and (2) using Monte Carlo algorithms, relying on Markov chain Monte Carlo (MCMC) and importance sampling, respectively. For background information regarding MCMC and importance sampling, we refer the reader to \citet{brooks2011handbook} and \citet{tokdar2010importance}.

\subsection{Computing the Posterior for $\Theta$}\label{MCMC}

We consider an augmented version of the posterior that instantiates the auxiliary variables $Z = (Z^s)_{s \in \jesa}$ and $U = (u^s)_{s \in \jesa}$. We use $\mathcal{L}(\Theta, Z, U)$ to denote the augmented likelihood
\begin{align}\label{likelihood}
\mathcal{L}\left(\Theta, Z, U \right) &= \frac{1}{\Gamma(q)^{|\jesa| |\mathcal{A}|}} \prod_{s \in \jesa}  \frac{u^{N_s - 1}}{\Gamma(N_s)}  \prod_{a \in \mathcal{A}} \frac{ \Gamma(q + C_a^s) \left(\common_{a} \contact^s_{a}\right)^{C^s_a}}{\left(u^s \common_{a}\contact^s_{a} + 1 \right) ^{q + C^s_a}} \prod_{n=1}^{N_s} \kernel( \Delta^s_n),
\end{align}
where $\Delta^s_n$ and $C^s_a$ are defined as in (\ref{marginal2}). Our target is the posterior distribution $\Theta$, $U$, $Z$, with density $p(\Theta, U , Z| (x^s, \mathcal{C}^s)_{s \in \jesa})$ satisfying:
\begin{align}
p(\Theta, U, Z| (x^s, \mathcal{C}^s)_{s \in \jesa}) \propto \mathcal{L}\left(\Theta, Z, U \right) p(\Theta). \label{fullposterior}
\end{align}

Our MCMC algorithm consists of sequential updates of the parameters --- akin to Metropolis within Gibbs --- with most of the components being updated according to slice sampling \citep{neal2003slice, murray2010elliptical}. The updates are repeatedly performed in the following sequence:

\begin{itemize}
\item Each auxiliary variable $(u^s)_{s \in \jesa}$ is updated one-by-one using slice sampling. These updates can be performed in parallel.
\item The entire vector $\common$ is updated jointly using elliptical slice sampling.
\item Each entry in $(\psi_i)_{i=1,\ldots, 32}$ is updated one-by-one using slice sampling.
\item The parameter $q$ is updated using a slice sampler.
\item Each entry in $p^h$ then $p^v$ is updated one-by-one using slice sampling.
\item Each auxiliary variable $(z^s_{n})$ is updated one-by-one by Gibbs sampling.
\end{itemize}

The details and conditional distributions for these updates are available in the Appendix (\S\ref{MCMCprops}). This algorithm provides a sequence of draws of $\Theta$ from its posterior that can be used to approximate posterior expectations. Notably, we can use these to approximate the posterior marginal probability of a configuration of accidentals (Task 2) as we now detail in $\S\ref{IS}$.

\subsection{Computing Marginal Densities via Importance Sampling}\label{IS}

A natural metric for assessing the performance of our model is to split $\jesa$ into a training set $T$ and test set $T'$, then evaluate the held out density of the accidental locations on each shoe in $T'$ (given $T$). Doing this requires computing
\begin{align}
p(x^\tau \mid \mathcal{C}^{\tau}, T) &= \mathbb{E}_{\Theta}\left(p(x^\tau| \mathcal{C}^{\tau}, \Theta) \mid (x^s, \mathcal{C}^s)_{s \in T}\right) \label{marglik}
\end{align}
for each $\tau \in T'$, where $p(\cdot \mid \mathcal{C}^{\tau}, T)$ denotes the posterior density. Here, $\mathbb{E}_{\Theta}(\cdot | (x^s, \mathcal{C}^s)_{s \in T})$ denotes the expected value under the posterior of $\Theta$ given the contact surfaces and accidentals in $T$. Note that the nested integrals in the expression in (\ref{marglik}) can be separated into an outer integral and an inner integral. The outer integral is the posterior expectation over the global parameters $\Theta$ and can be approximated using MCMC draws as described above. The inner integral --- computed for each posterior draw --- is over the local auxiliary variables $u^{\tau}$ and $Z^{\tau}$ as shown in (\ref{marginal2}). 

We approximate this integral using importance sampling. Specifically, given a draw of $\Theta$, we define an importance distribution given by
\begin{align}
u \mid \Theta, N_s, \mathcal{C}^{\tau} &\sim \text{Gamma}\left(N_s,  q \sum_{a \in \mathcal{A}} \common_a \contact^\tau_{a}\right) \label{importanceu}\\ 
\mathbb{P}(Z_n = x_n^{\tau} + a \mid \Theta, x^{\tau}_n ) & = \frac{\common_{a + x_n^{\tau}} \contact^{\tau}_{a +  x_n^{\tau}} \kernel(a)}{\sum_{b \in B}\common_{b + x_n^{\tau}} \contact^{\tau}_{b +  x_n^{\tau}} \kernel(b)} \label{importancez}
\end{align}
where $B = \{-3,\ldots, 3\}^2$ and $a \in B$ for all $n \in\{ 1, \ldots, N_s\}$. After drawing $M > 0$ importance samples  $u_1, \ldots, u_M \in \mathbb{R}_+$ by (\ref{importanceu}) and $Z^{1}, \ldots, Z^{M} \in \mathcal{A}^{N_{\tau}}$ by (\ref{importancez}), the inner integral can be approximated as
\begin{align*}
p\left(x^\tau| \mathcal{C}^\tau, \Theta \right) &= \frac{\prod_{n=1}^{N_{\tau}} \left( \sum_{b \in B}\common_{b + x_n^{\tau}} \contact^{\tau}_{b +  x_n^{\tau}} \kernel(b) \right)}{\Gamma(q)^{|\mathcal{A}|} \left( q \sum_{a \in \mathcal{A}} \common_a \contact^{\tau}_a \right)^{N_\tau}} \sum_{m=1}^M \frac{ \exp{(u^m q \sum_{a \in \mathcal{A}} \common_a \phi^{\tau}_a)}}{ \prod_{a \in \mathcal{A}} \frac{\left(u^m \common_a \contact_a^\tau + 1 \right)^{q + C^m_a}}{\Gamma(C^m_a + q)}},
\end{align*}
where $C^M_a$ denotes the number of times $a \in \mathcal{A}$ occurs as an entry in $Z^M$.

Thus, using one importance sample ($M=1$) for each MCMC draw $\Theta^{\ell} = (\contact^{\ell}, \common^{\ell}, q^{\ell}, (p^h)^{\ell}, (p^v)^{\ell})$ yields the approximation
\begin{align}
p(x^\tau \mid \mathcal{C}^{\tau}, T)  \approx \sum_{\ell=1}^L \frac{\prod_{n=1}^{N_{\tau}} \left( \sum_{b \in B}\common_{b + x_n^{\tau}} (\contact^\ell)^{\tau}_{b +  x_n^{\tau}} \kernel^\ell(b) \right)}{ L \Gamma(q^\ell)^{|\mathcal{A}|} \left( q^\ell \sum_{a \in \mathcal{A}} \common_a (\contact^\ell){\tau}_a \right)^{N_\tau}} \frac{ \exp{(u^\ell q^\ell \sum_{a \in \mathcal{A}} \common^\ell_a (\contact^\ell)^{\tau}_a)}}{ \prod_{a \in \mathcal{A}} \frac{\left(u^\ell \common^\ell_a (\contact^\ell)_a^\tau + 1 \right)^{q^\ell + C^\ell_a}}{\Gamma(C^\ell_a + q^\ell)}},
\end{align}
where $L$ is the total number of MCMC draws and the $(u^{\ell}, Z^{\ell})_{1 \leq \ell \leq L}$ are each drawn according to the respective importance distribution for $\Theta^{\ell}$. Detailed derivations and discussion of this strategy are available the Appendix (\S\ref{fullIS}).

\section{Comparisons to Competitors and Summary of Fit}\label{comparisons}

\subsection{Comparison to Competitors}\label{competitors}

To demonstrate that efficacy of our model, we compare its performance to three competitor models. The first two models we consider -- the uniform model of \cite{stone2006footwear} and the kernel density estimator of \cite{yekutieli2012expert} -- rely on fitting a single fixed density $\Lambda$ for all shoes. Recall from \S \ref{problem_form} that the kernel density estimator does not make use of contact surface information when estimating $\Lambda$, and that the uniform model does not rely on any data at all. 

For this reason, we introduce a third competitor called the \emph{contact model}. In the contact model, each $G_{\mathcal{C}^s}$ is defined as a point mass at $\Lambda_{\mathcal{C}^s}$ with
\begin{align}\label{contactmodel}
\Lambda_{\mathcal{C}^s}(a) \propto \exp(\alpha_{r^s_{a}}).
\end{align}
Here, $\alpha \in \mathbb{R}^{32}$ are shared amongst all of $G$, similar to $\contact$ with $r^s_a$ following the same set-up as defined as in (\ref{rsa}). The parameters $\alpha$ are straightforward to infer using maximum likelihood (fixing $\alpha_1 = 1$ to obtain identifiability).

We fit our model and the three competitor models to four test/train splits of the JESA data, with each training set consisting of 336 randomly selected shoes. The remaining 50 serve as the test set. For our model, the posterior was computed by running the MCMC algorithm outlined in Section~\ref{MCMC} for 30000 full sweeps and discarding the first 10000 iterations as warm-up. 

\begin{figure}[ht]
  \centering
  \input{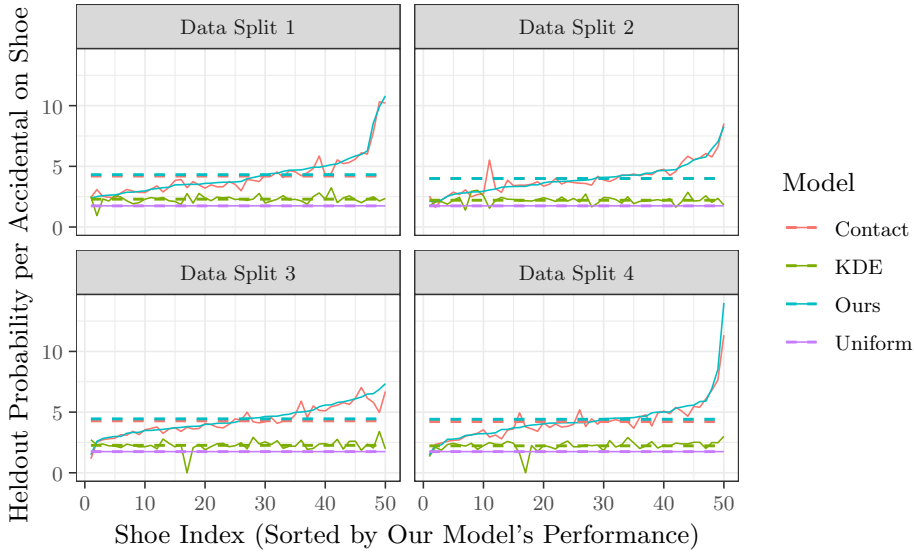}
  \caption{Comparison of the performance of four models: the contact model (red) the kernel density estimate (green), our model (blue) and the uniform model (purple) on 50 held out shoes across four data splits. The solid lines depict the metric given in (\ref{plotmetric}) for each of 50 shoes (sorted by our model's performance). The dotted lines depict the mean for each model.} \label{performance}
\end{figure}

 Let $T$ denote a training set and $T'$ denote the test set. As a metric of performance, we used our importance sampling technique to evaluate the held-out density of the accidental locations $x^{\tau}$ on each shoe $\tau \in T'$ given $T$. Figure~\ref{performance} depicts the held-out likelihood per accidental on each held-out shoe for each of the four models fit to each of the four splits. Specifically,
\begin{align}\label{plotmetric}
20000 \times p(x^\tau \mid \mathcal{C}^{\tau}, T)^{1/N_{\tau}} 
\end{align}
is reported for each $\tau \in T'$. The scaling by 20000 is performed for readability of the $y$-axis (it is equivalent to transforming $\mathcal{A}$ to the unit square) and the $N_{\tau}$th root is taken to facilitate comparison of average performance on shoes with different numbers of accidentals. This metric is equivalent to comparing the per-accidental average log loss of each shoe. The held-out shoes were sorted according to our model's performance for each of the four splits. Note that for the uniform model, only those atoms in $\mathcal{A}$ were given positive density, hence the constant density of 1.743 rather than 1.

It is evident from Figure~\ref{performance} that the two models that account for contact surface (our model and the simple contact surface model (\ref{contactmodel})) vastly outperform the two that do not. Notably, the kernel density estimator assigns 0 density to a shoe in splits 3 and 4, showing an alarming lack of robustness. The performance of our model and the contact model tend to track together across shoes, suggesting that the incorporation of the contact surface is the major driver of both models' success.

We also checked whether the other components of the model ($w$, $\kappa$, and $\noise$) contribute positively to the model's performance. We fit an additional five variants of our model to the training data and summarized their results in Table~\ref{resultstab}, along with the performance of the four original competitors. The variant models are defined as follows. ``Without scores'' refers to our model with all $\noise^s_a$ variables are fixed at one, ``without kernel'' refers to our model but without $\kernel$ smoothing, ``without scores and kernel'' excludes both $\noise^s_a$ and $\kernel$, ``without $w$'' fixes $\common_E = 1$, and ``without $\phi$'' fixes all $\phi$ at 1. Posterior computation for all variant models were performed using appropriate analogs of the MCMC algorithm given in \S \ref{MCMC}.

For each model and test set $T'$, Table~\ref{resultstab} reports the geometric mean of (\ref{plotmetric}) across all held-out shoes, i.e.
\begin{align}\label{plotmetric2}
20000 \times \left(\prod_{\tau \in T'} p(x^\tau \mid \mathcal{C}^{\tau}, T)^{1/N_{\tau}}\right)^{1/|T'|}.
\end{align}
This metric is equivalent to the mean per-accidental log loss across shoes. 

\begin{table}[]
\begin{tabular}{l|l|l|l|l|l|}
\cline{2-6}
                                                                                                        & Method     & Split 1 & Split 2 & Split 3 & Split 4 \\ \hline \hline 
                                                                                                        
\multicolumn{1}{|l|}{\multirow{2}{*}{\begin{tabular}[c]{@{}l@{}}Existing\\ Models\end{tabular}}}        & Uniform \citep{stone2006footwear}    & 1.743   & 1.743   & 1.743   & 1.743   \\ \cline{2-6} 
\multicolumn{1}{|l|}{}                                                                                  & KDE  \citep{yekutieli2012expert}       & 2.266    & 2.182  & 0.000   & 0.000   \\ \hline \hline
\multicolumn{1}{|l|}{Other Models}                                                                      & Contact & 3.954   & 3.823   & 4.106  & 3.995       \\ \hline \hline
\multicolumn{1}{|l|}{\multirow{6}{*}{\begin{tabular}[c]{@{}l@{}}Our Model \\
and variants \end{tabular}}} & Full        & \textbf{4.060}   & 3.832   & \textbf{4.272}   & \textbf{4.144}   \\ \cline{2-6} 
\multicolumn{1}{|l|}{}                                                                                  &  without scores      & 4.052   & 3.831   & 4.260  & 4.131   \\ \cline{2-6} 
\multicolumn{1}{|l|}{}                                                                                  & without kernel     & 4.041   & 3.794   & 4.244   & 4.081  \\ \cline{2-6} 
\multicolumn{1}{|l|}{}                                                                                  & without scores and kernel  & 4.039  & 3.791  & 4.238   & 4.072   \\ \cline{2-6} 
\multicolumn{1}{|l|}{}                                                                                  & without $w$       & 3.981   & \textbf{3.860}   & 4.131   & 4.070   \\ \cline{2-6} 
\multicolumn{1}{|l|}{}                                                                                  & without $\phi$ & 2.217   & 2.124   & 2.187   & 2.144  \\ \hline 

\end{tabular}
\caption {The mean predictive performance (measured by (\ref{plotmetric2})) of our model, five variants on our model, and three competitor models. The best performing result is bolded for each split.} 
\label{resultstab} 
\end{table}

Table~\ref{resultstab} demonstrates that our full model outperforms all competitors and variants on Splits 1, 3, and 4, being edged out only by ``without $w$'' on Split 2.  Nearly all variants perform close to comparably to the full model; the notable exception is ``without $\phi$''. It performs far worse, highlighting the importance of the contact surface. The persisting decrease in performance of the other variants across splits indicates that each component provides a small gain, and is worth keeping in the model. 

Note that the superior performance of ``without $\common$'' in Split 2 is explained by the presence of an atypical shoe in the test set.  
It possesses only two accidentals, both of which are located at the left side of the heel. As illustrated in Figure~\ref{predictives}($\common$), $\common$ is small towards the heel, especially on the lefthand side. Consequently, including $w$ leads to far lower predictive posterior probability for this particular shoe. Excluding this shoe from the test set 2 results in the full model regaining its spot as the top performer.

\subsection{Summary of Inferred Model Parameters}\label{params}

\begin{figure}[!h]
\centering
\input{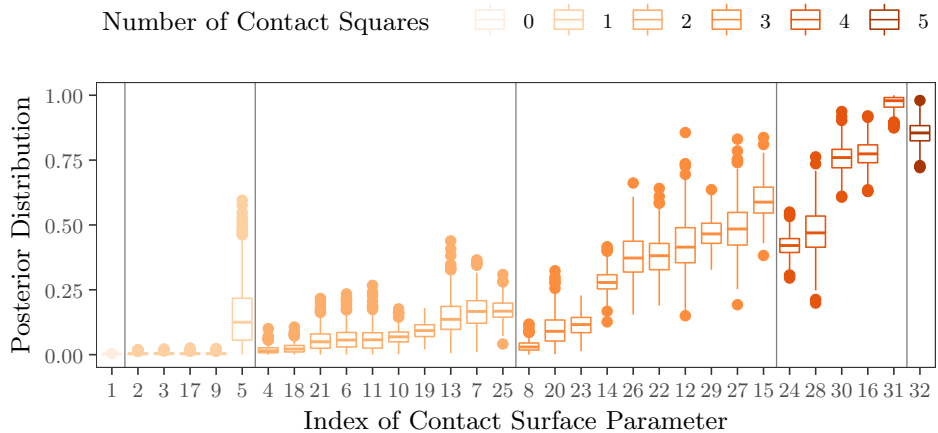}
\caption{Posterior distribution boxplots of the parameters of the 32 possible shapes (listed in Figure~\ref{phi}). Boxplot color indicates with the amount of contact surface present in each, with vertical lines partitioning the levels.}\label{contact_posterior}
\end{figure}

\begin{figure}[!h]
  \centering
  \input{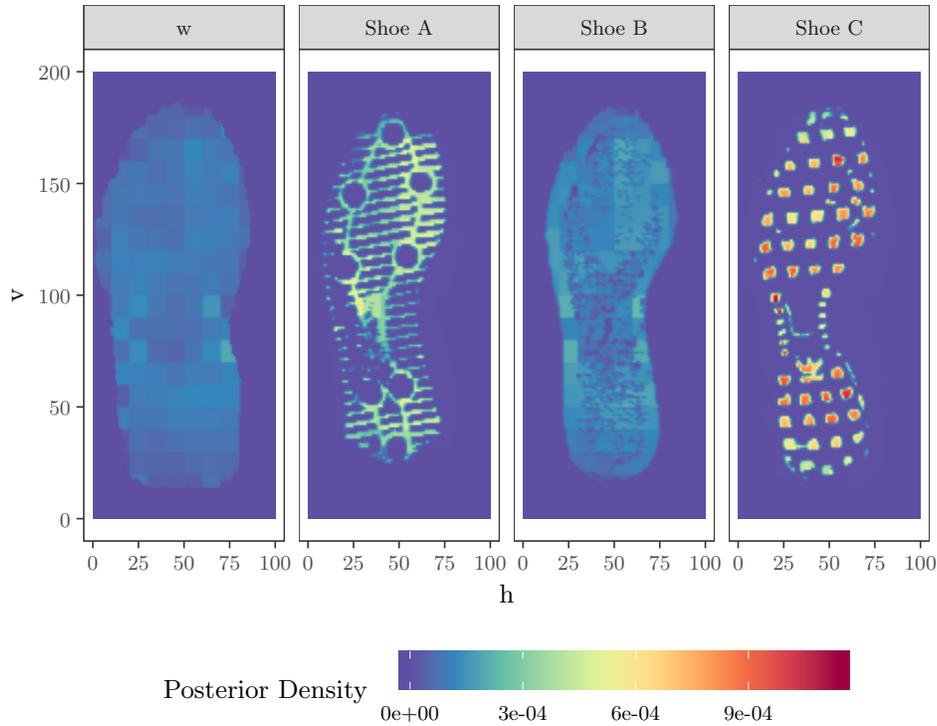}
  \caption{Panels $w$, Shoe A, Shoe B, and Shoe C demonstrate the posterior predictive distribution of accidental locations for four contact surfaces. Panel $w$ is synthetic (entirely contact surface). Shoe A corresponds to the shoe shown in Figure~1. Shoes B and C are other contact surfaces  from JESA.} \label{predictives}
\end{figure}

To investigate our fitted model, we consider the posterior of $\Theta$ from Split 1 in \S \ref{competitors}. Components of the posterior distribution are summarized in Figures~\ref{kernel}, \ref{contact_posterior}, and \ref{predictives}.

Figure~\ref{kernel} summarizes the posterior fit for the kernel $\kernel$. Figure~\ref{kernela} uses boxplots to demonstrate the posterior distribution of both $\kappa^h$ and $\kappa^v$, arranged symmetrically to facilitate visualization of the kernel. For both $h$ and $v$, the kernel's mass is mostly concentrated on its mode and immediate neighbours. The smoothing is also more diffuse in the horizontal direction that the vertical direction, suggesting that the accidental distributions are smoother in the horizontal direction that vertical direction. Figure~\ref{kernelc} demonstrates the composition of the vertical and horizontal kernel into the bivariate kernel. 

Figure~\ref{contact_posterior} displays the marginal posterior distributions of each $\phi_1, \ldots,\phi_{32}$ using boxplots. Here, the larger the associated posterior value, the more likely an accidental is to occur nearby contact surface taking on the shape. There is a stark difference in accidental proclivity between gridpoints surrounded mostly by contact surface (shapes 32, 31, 30, 28, 24, 16 as depicted in Figure~\ref{phia}) and those with little contact surface present (shapes 1, 2, 3, 5, 9, 17). This difference supports the intuition among shoeprint examiners that regions which rarely make contact with the ground are typically less likely to accumulate accidentals. Also notable is the discrepancy between different shapes containing the same amount of contact surface. For example, accidentals appear to be nearly to twice as likely to be associated with gridpoints exhibiting shape 31 than those exhibiting shape 24, even though both shapes consist of 4 of 5 possible contact components. This inference suggests the shape of the contact surface --- and not just the amount of contact surface --- also plays a role in a region's likelihood of being marked with an accidental. However, we caution against over-interpreting such differences due to $\phi$ being just one component of the larger model.

Figure~\ref{predictives} illustrates the posterior predictive distribution of an accidental for four separate contact surfaces. The first panel is a synthetic, consisting entirely of contact surface in order to demonstrate $w$. The inward facing side of the toe tends to  to more exhibit accidentals than the in outward facing portion, and the front of the heel tends to exhibit more accidentals than the rear of the heel. A depiction of the fit and uncertainty of the raw $w$ parameter is available in the Appendix (Figure~\ref{Jstand}).

The second through fourth panels of Figure~\ref{predictives} (Shoe A, Shoe B, Shoe C) demonstrate the posterior mean of $\Lambda^s$ for three example contact surfaces in JESA. The difference in the magnitude of the density between Shoe B and Shoe C demonstrates that the density associated with a particular location is heavily contingent on the total amount of contact surface present for the shoe; because shoe C demonstrates relatively little contact surface, the density is much higher in locations where contact surface is present.


\section{Discussion} \label{discussion} 

In this work, we made progress on a problem put forth by the President's Council of Advisors in Science and Technology \citep{president2016report}. Namely, we formalized the problem of modeling accidental distributions for random match probabilities, developed a modeling framework for the spatial distribution of accidentals on shoe soles, fit our hierarchical Bayesian model to real data within a Bayesian nonparametric setting to pool information across a variety of shoes, and demonstrated that our model vastly outperforms existing models in the literature on a held-out data task.

A key takeaway from this endeavor was the importance of explicitly incorporating the contact surface when modeling accidental distributions. We were the first to do so, and it resulted in a major improvement over the traditional models. We took care to develop our model hierarchically, allowing for the pooling information across shoes of different types to capture commonalities in how the contact surface influences accidental distributions. As data sources grow and new data collection efforts are undertaken \citep{CSAFE2019longitudinal}, we anticipate the opportunity for more sophisticated models to better capture the relationship between contact surface and accidentals. 

Along these lines, a natural extension of our model would be to allow the $\common$, $\contact$, and $q$ parameters to differ across shoes according to a nonparametric mixture model.  Another possibility would be to extend the model to a spatiotemporal setting, using the temporal data being collected by \cite{CSAFE2019longitudinal} to model how accidentals accumulate over time.

A possible limitation of our model stems from treating the contact surface parameter $\contact$ and spatial location parameter $\common$ separately. It is plausible that a shoe's intensity would involve dependence between the contact surface and the spatial location. For instance, accidentals could be more likely to occur in high contact areas when on the toe, but more likely to occur in low contact areas when on the heel. In such instances, a model including an interaction effect would outperform our current model.

Another issue we briefly touched on without addressing was the open problem of formally defining when two impressions ``match '' ($x^s \equiv x^y$). Given a similarity metric defining when $x^s \equiv x^y$, our model is tailored to computing the RMP. Draws from the posterior distribution of $x^s| \mathcal{C}^s$ can serve as a surrogate for sampling from $\pop_{\mathcal{C}^y}$ in (\ref{rmpv}), providing a straightforward Monte Carlo strategy for evaluating the RMP. It is worth noting that although our exposition focused on RMPs, our approach is equally applicable to calculating other related summaries of uncertainty, such as likelihood ratios or Bayes factors \citep{evett1998bayesian}.

Finally, we would like to highlight uses of our model outside of direct evaluation of random match probabilities. Recently, the National Institute for Standards in Technology has started development of a multipurpose software tool for forensic footwear examiners \citep{herman2016measurements}. One of the tools in development is ShoeGuli, a program for developing synthetic footwear impressions complete with accidentals. As our framework results in an accurate generative model, it is a natural choice for simulating accidental patterns.

\section*{Acknowledgements}
Thank you very much to Yaron Shor and Sarena Wiesner of the Israeli Police Department and Yoram Yekutieli of Hadassah Academic College for sharing and explaining the details of the JESA database. Thank you to our colleagues in Center for Statistics and Applications in Forensic Evidence and Martin Herman, Hari Iyer, Steven Lund, and others at NIST for advice and feedback. Special thanks to Steve Fienberg for his encouragement and involvement in the early stages of this project.
\bibliography{shoesref}
\bibliographystyle{unsrtnat}
\newpage
\section{Appendix}

\subsection{Discretization and Kernel Choice} \label{discretization}

Recall from \S \ref{data} that the contact surface variables are defined on a discrete 100 by 200 equally-spaced grid over $[0,100] \times [0,200]$. For practicality, we restrict our model for $\Lambda_s$ to match the resolution of this grid, discretizing our kernel $\kernel$ to be piece-wise constant over each gridpoint. Theoretically, such a choice restricts the model's flexibility at resolutions smaller than that of the grid. However, we do not expect these resolutions to be relevant to RMP calculations --- any such effect will be dominated by the noise in the observed accidental locations for crime scene prints. Attempting to model any structure at such a resolution would amount to overfitting.

In addition to preventing overfitting and facilitating interpretation, the discretized kernel provides computational and modeling advantages. 

Computationally, the discretization eliminates the need to keeping track of each real valued accidental locations $x^s_n \in [0,100] \times [0,200]$. Instead, we need only store the discrete gridpoint values in $\mathcal{A} = \left\{1,\ldots, 100\right\} \times \left\{1,\ldots, 200\right\}$. Similarly, we can directly store a $\Lambda_s$ as a vector of real values and sampling accidental locations from it is equivalent to simply drawing from a multinomial. The grid $\mathcal{A}$ also provides a natural resolution with which to visualize $\Lambda_s$. In addition to these computational conveniences, discretization also provides computational speed-ups of the Bayesian inference procedure.

The speed-up is best illustrated by comparing to the standard unimodal symmetric kernel choice in Bayesian nonparametrics --- the Gaussian density. If we were to replace the $\kernel$ in the model with a bivariate Gaussian density, every kernel function would have positive density over the entire $[0,100] \times [0,200]$ spatial domain. Consequently, each auxiliary variable $Z^s_{n}$ could take on any of $|\mathcal{A}|$ values instead of the 49 associated with our discrete kernel. Exploring the large space of possible $Z^s_n$ would lead to additional computational burden for both our MCMC and importance sampling algorithms. Eventually we would need to evaluate $|\mathcal{A}|$ Gaussian densities for each data point.

Moreover, the Gaussian densities would need to be truncated at values outside the $[0,100] \times [0,200]$, requiring the computation of normalization constants for any gridpoints close enough to the boundary. If the goal was to infer the bandwidth of the kernel as part of the MCMC procedure, these normalization constants would need to be repeatedly recalculated for each updated bandwidth.

From a modeling perspective, the discrete kernel model also makes more sense than the Gaussian kernel in the context of our data. For many shoes in JESA (e.g. shown in Figure~\ref{contact3}), the contact surface drops off steeply and remains zero for large portions of the space, leaving no opportunity for accidentals in these regions. The discrete kernel can accommodate this behavior; its redistribution of the density is restricted to be very local, preventing it from directing density toward these impossible regions. In contrast, the smooth decay of a Gaussian kernel forces it to assign at least some density from each kernel to the entire space, regardless of the contact surface in that area. To limit the density wastage around these steep drop-offs, inference of the Gaussian kernel would promote a very small bandwidth, thus limiting the amount of possible smoothing. Our discrete kernel is better equipped to deal with such a problem, its flexible parameterization allows it to distribute most of the density over nearby gridpoints.

\subsection{Additional Related Work}\label{relatedwork2}

The NCoRM framework represents one of many models for collections dependent probability distributions that use normalized random measures. The prototypical normalized completely random measure is the Dirichlet process \citep{ferguson1973bayesian} which serves as a building block for much of the literature. Within the spatial statistics literature, Dirichlet process mixture models were first applied by \cite{gelfand2005bayesian} in the context of modeling random functions in space. They have also been applied to model intensities for spatial point processes (e.g. \cite{kottas2007bayesian, taddy2010autoregressive, jewell2015atomic}). Popular approaches for modeling vectors of dependent probability distributions include the dependent Dirichlet process \citep{maceachern2000dependent}, the hierarchical Dirichlet process \citep{teh2005sharing}, and the nested Dirichlet process \citep{rodriguez2008nested}. Non-Dirichlet process-based techniques include \citep{chen2013dependent, foti2012slice, lijoi2014bayesian}. 

Much of the literature pertaining to vectors of probability measures assumes that the vectors are exchangeable, with the dependent Dirichlet process \citep{maceachern2000dependent} and the kernel stick-breaking process \citep{dunson2008kernel} comprising two notable exceptions. Other recent work pertaining to the modeling vectors of non-exchangeable probability distributions was surveyed in \citet{foti2015survey}. However, we found that the existing literature lacked the tools to incorporate our desired dependence structure for the shoes in JESA, which prompted us to extend the NCoRM framework.

Contrasting with completely random measure-based techniques, another frequently used tool for modeling spatial point processes is the log-Gaussian Cox process \citep{moller1998log, adams2009tractable}. The log-Gaussian Cox process is able to capture more sophisticated spatial dependencies by explicitly modeling the log intensity as a draw from a Gaussian process, with the kernel of this process prescribing the spatial correlation structure. We draw on this work by using a log-Gaussian prior on $\common_E$ (a finite resolution log-Gaussian process).

\subsection{Details of Parameterization of $\common$} \label{commondetails}

\begin{figure}[h]
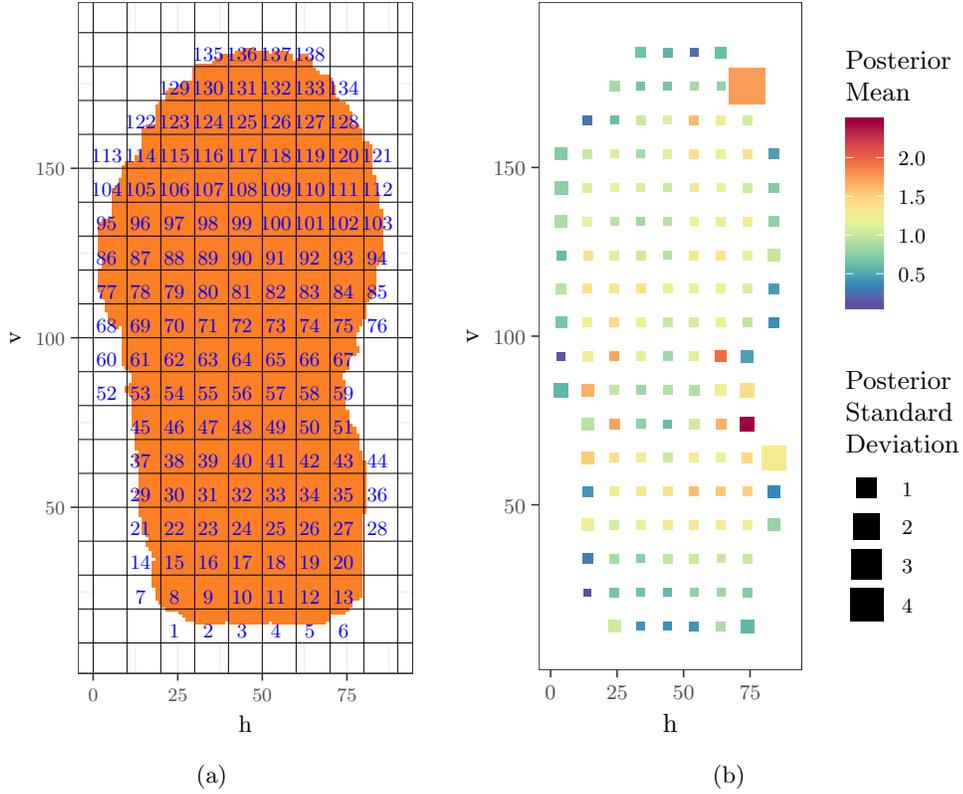

\centering
  \subfloat[]{ \input{Jgrid.tex} \label{Jgrid}}
    \subfloat[]{\input{Jstandard_deviation.tex} \label{Jstand}}
\caption{(a) displays the 20000 gridpoints $a \in \mathcal{A}$, colored white if we fix $\common_a  = 0$, orange otherwise. The black lines partition $\mathcal{A}$ into the coarser $10 \times 10$ grid associated with $\common_E$, each nonzero grid region contains a blue number indicating the index in $\common_E$ to which it corresponds. (b) summarized the posterior distribution of $\common_E$ as a square corresponding to each entry in $\common_E$. As per the legend, the color of the square indicates its posterior mean and the size of the square indicates is posterior standard deviation.}
\end{figure}

Because inferring 20000 unique entries $\common$ represents a large computational burden, it is helpful to reduce its dimension by parametrizing it as piece-wise constant over a coarser region. We define these regions, illustrated in Figure~\ref{Jgrid}, using two criteria. First, we reduce of the resolution from the original $200 \times 100$ grid to a $20 \times 10$ grid of unique values, with each new region now corresponding to 100 of the original grid points. Second, it is evident from Figures~\ref{aggregatedcontacts} and \ref{accidentallocations} that a sizable proportion of $\mathcal{A}$ --- specifically the gridpoints at the sides and extremities of the bounding box --- have no practical probability of being marked by an accidental. We choose to force their respective $\common_a$'s to be 0 in the prior, essentially omitting them from analysis. 

After this restriction, we use the remaining grid regions that have at least one positive atom to define our 138 distinct regions. We use $\common_E \in \mathbb{R}_+^{138}$ to denote the vector of unique values assigned to each of these regions, assigning it a lognormal prior. The prior mean for $\log(\common_{E})$ is fixed at 0. The precision $\Sigma$ is fixed such that each diagonal entry is 1. Off-diagonal entires are 0 for non-adjacent regions, 0.2 otherwise. The full mapping between the entries in $\mathcal{A}$ and the indices of $\common_E$ is displayed in Figure~\ref{Jgrid} with the nonzero gridpoints depicted as orange pixels. Throughout the article, $\mathcal{A}$ refers to the subset of $\{1, \ldots, 100\} \times \{1, \ldots, 200 \}$ that correspond to the nonzero gridpoints. 

\subsection{Details of Marginalization of $\noise^s$} \label{marginalization}
Recall that $\Gamma(\cdot)$ denotes the gamma function. Let $\Delta_n^s \in \{-3,3\}^2$ be shorthand for $Z_n^s - x_n^s$, and let
\begin{align}
\zeta^s = \left\{Z^s : \Delta_n^s \in \{-3,3\}^2, n \in \left\{1,\ldots, N_s\right\} \right\}
\end{align}
denote the set of possible values for $Z^s$. After introducing $Z^s$, the marginal density can be re-expressed as
\begin{align}\label{our_rmp2}
p(x^s| \mathcal{C}^s; \Theta) &=  \sum_{Z^s \in \zeta^s } \left( \int  \prod_{n=1}^{N_s} \kernel(\Delta_n^s) \frac{\common_{Z^s_n} \noise_{Z^s_n}^{s} \contact^s_{Z^s_n}}{\sum_{a \in \mathcal{A}} \common_{a} \noise_{a}^{s} \contact^s_{a} }  \text{d}\rho(\noise^s)\right) \\
                                                                 &= \sum_{Z^s \in \zeta^s } \left(  \prod_{n=1}^{N_s} \kernel(\Delta_n^s) \int \frac{\prod_{a \in \mathcal{A}} \left(\common_{a} \noise_{a}^{s} \contact^s_{a}\right)^{C^s_a}}{\left(\sum_{a \in \mathcal{A}} \common_{a} \noise_{a}^{s} \contact^s_{a} \right) ^{N_s}}  \text{d}\rho(\noise^s)\right).
\end{align}

Let $\text{Ga}( \cdot | \alpha, \beta)$ denote the probability density function of a Gamma distribution with shape $\alpha$ and rate $\beta$. Recall that
\begin{align}
u^s \sim \text{Gamma}\left( n_s, \sum_{a \in \mathcal{A}} \common_{a} \noise_{a}^{s} \contact^s_{a}\right).
\end{align}
Incorporating the density of $u^s$ into (\ref{our_rmp2}) allows to analytically marginalize the $\noise^s_a \sim \text{Gamma}(q,1)$ to derive a simpler expression for $ \text{rmp}_V(x^s| \mathcal{C}^s; \Theta)$.
\begin{align*}
 & \sum_{Z^s \in \zeta^s } \left( \prod_{n=1}^{N_s} \kernel(\Delta_n^s) \int \frac{\prod_{a \in \mathcal{A}} \left(\common_{a} \noise_{a}^{s} \contact^s_{a}\right)^{C^s_a}}{\left(\sum_{a \in \mathcal{A}} \common_{a} \noise_{a}^{s} \contact^s_{a} \right) ^{N_s}}  \text{d}\rho(\noise^s)\right)\\
 &= \sum_{Z^s \in \zeta^s } \left(\prod_{n=1}^{N_s} \kernel(\Delta_n^s) \int \frac{\prod_{a \in \mathcal{A}} \left(\common_{a} \noise_{a}^{s} \contact^s_{a}\right)^{C^s_a}}{\left(\sum_{a \in \mathcal{A}} \common_{a} \noise_{a}^{s} \contact^s_{a} \right) ^{N_s}} \int \text{Ga} \left( u^s \mid n_s, \sum_{a \in \mathcal{A}} \common_{a} \noise_{a}^{s} \contact^s_{a} \right) \text{d}u^s \text{d}\rho(\noise^s)\right)   \\
 &=   \sum_{Z^s \in \zeta^s } \left( \prod_{n=1}^{N_s} \kernel(\Delta_n^s) \int \int \text{Ga} \left( u^s \mid n_s, \sum_{a \in \mathcal{A}} \common_{a} \noise_{a}^{s} \contact^s_{a} \right) \frac{\prod_{a \in \mathcal{A}} \text{Ga}\left(\noise^s_a | q,1 \right) \left(\common_{a} \noise_{a}^{s} \contact^s_{a}\right)^{C^s_a}}{\left(\sum_{a \in \mathcal{A}} \common_{a} \noise_{a}^{s} \contact^s_{a} \right) ^{N_s}} \text{d}\noise^s \text{d}u^s\right) \\
                                                                 &= \sum_{Z^s \in \zeta^s } \left(\prod_{n=1}^{N_s} \kernel(\Delta_n^s) \int \int \frac{\prod_{a \in \mathcal{A}} \left(\common_{a} \noise_{a}^{s} \contact^s_{a}\right)^{C^s_a}}{\left(\sum_{a \in \mathcal{A}} \common_{a} \noise_{a}^{s} \contact^s_{a} \right) ^{N_s}} \text{Ga} \left( u^s \mid n_s, \sum_{a \in \mathcal{A}} \common_{a} \noise_{a}^{s} \contact^s_{a} \right) \prod_{a \in \mathcal{A}} \text{Ga}\left(\noise^s_a | q,1 \right) \text{d}\noise^s \text{d}u^s\right)\\
                                                                 &= \sum_{Z^s \in \zeta^s } \left(\prod_{n=1}^{N_s} \kernel(\Delta_n^s) \int \int  \prod_{a \in \mathcal{A}} \frac{\left(\common_{a} \noise_{a}^{s} \contact^s_{a}\right)^{C^s_a}}{ \Gamma(q) (\noise_{a}^{s})^{1-q}} \frac{(u^s)^{N_s - 1}}{\Gamma(N_s)} \exp{\left(- \sum_{a \in \mathcal{A}} (\common_a \contact^s_a u_s + 1) \noise_a^s \right)} \text{d}\noise^s \text{d}u^s\right)\\
                                                                 &= \sum_{Z^s \in \zeta^s }\left(\prod_{n=1}^{N_s} \kernel(\Delta_n^s) \int\frac{(u^s)^{N_s - 1}}{\Gamma(N_s)} \prod_{a \in \mathcal{A}} \frac{\left(\common_{a} \contact^s_{a}\right)^{C^s_a}}{ \Gamma(q)}\left( \int_{0}^{\infty} \left( \noise_{a}^{s}\right)^{C^s_a + q - 1}  \exp{\left(- (\common_a \contact^s_a u_s + 1) \noise_a^s \right)} \text{d}\noise_a^s \right) \text{d}u^s\right)\\
                                                              &= \sum_{Z^s \in \zeta^s } \left(\prod_{n=1}^{N_s} \kernel(\Delta_n^s) \int\frac{(u^s)^{N_s - 1}}{\Gamma(N_s)} \prod_{a \in \mathcal{A}} \frac{\left(\common_{a} \contact^s_{a}\right)^{C^s_a}}{ \Gamma(q)}  \frac{\Gamma(N_s + q)}{\left(\common_{a} \contact^s_{a} u^s + 1 \right)^{q + C_a^s}} \text{d}u^s\right)\\
                                                              &= \frac{1}{\Gamma(q)^{|\mathcal{A}|}} \int_{0}^{\infty} \frac{u^{N_s - 1}}{\Gamma(N_s)}\sum_{Z^s \in \zeta^s }  \left( \prod_{n=1}^{N_s} \kernel(\Delta_n^s)\prod_{a \in \mathcal{A}} \frac{ \Gamma(q + C_a^s) \left(\common_{a} \contact^s_{a}\right)^{C_a}}{\left(u^s \common_{a}\contact^s_{a} + 1 \right) ^{q + C^s_a}} \right) \text{d}u^s\\
                                                              &= \frac{1}{\Gamma(q)^{|\mathcal{A}|}} \int_{0}^{\infty} \frac{u^{N_s - 1}}{\Gamma(N_s)}  \mathbb{E}_{Z^s}\left(\prod_{a \in \mathcal{A}} \frac{ \Gamma(q + C_a^s) \left(\common_{a} \contact^s_{a}\right)^{C_a}}{\left(u^s \common_{a}\contact^s_{a} + 1 \right) ^{q + C^s_a}} \right) \text{d}u^s
\end{align*}
where $\mathbb{E}_{Z^s}$ denotes the expectation with respect to the distribution of $Z^s$ given by (\ref{distz}).

\subsection{Details of MCMC Proposal Steps} \label{MCMCprops}

\subsubsection{MCMC update for $Z^s_n$}

Let $Z^s_{-n}$ denote $(Z^s_{i})_{i \neq n}$ and $C^s_{a, -n}$ denote $C^s_{a} - I(Z^s_n = a)$. We update each $Z^s_n$ using a Gibbs step, sampling from the conditional distribution of $Z^s_n \mid x^s, \mathcal{C}^s, Z^s_{-n}, u^s, \contact, \common, q, k$ given by
\begin{align}
\mathbb{P}(Z^s_n = z) \propto (q + C^s_{z, -n}) \frac{\common_{z} \phi^s_{z}}{u^s \common_{z} \phi^s_{z} + 1} k(x^s_n - z).
\end{align}

\subsubsection{MCMC update for $q$}

Let $\mathcal{D}_i = \sum_{s \in \jesa} \sum_{a \in \mathcal{A}} I(C_a^s = i)$ for non-negative integers $i$, and let $\mathcal{D} = (\mathcal{D}_i)_{0 \leq i \leq B}$ where $B$ is equal to the largest value of $i$ for which $\mathcal{D}_i > 0$. We update $q$ using a slice sampler on the condition distribution of
$q \mid Z, U, \contact, \common, (\mathcal{C}^s)_{s \in \jesa}, \mathcal{D}$ with density proportional to
\begin{align}
p(q) &\propto \frac{q \exp{(-2q)} \prod_{i=0}^B \Gamma(q + i)^{\mathcal{D}_i}}{\left(\prod_{s \in \jesa}\prod_{a \in \mathcal{A}} \left(u^s \common_a \phi^s_a + 1 \right)\right)^q}.
\end{align}
We use the stepping out method as described by \cite{neal2003slice}, with a step width of 0.2. Note that the computation of $\left(\prod_{s \in \jesa}\prod_{a \in \mathcal{A}} \left(u^s \common_a \phi^s_a + 1 \right)\right)$ can be recycled as the stepping out algorithm runs, and that the equality $\Gamma(q + i +1) = (q + i) \Gamma(q+i)$ can be exploited to speed-up the calculation of $\prod_{i=0}^B \Gamma(q + i)^{\mathcal{D}_i}$.

\subsubsection{MCMC update for $u^s$}
We update each $u^s$ using a slice sampler for the conditional distribution of
$u^s \mid Z^s, \contact, \common, q, (\mathcal{C}^s)_{s \in \jesa}$ with density proportional to 
\begin{align}
p(u) &\propto \frac{u^{N_s - 1}}{\prod_{a \in \mathcal{A}} \left(u \common_a \contact^s_a + 1 \right)^{q + C^s_a}}.
\end{align}
Our slice sampler uses the stepping out method as described by \cite{neal2003slice}, with a step width given by $20 \sqrt{N_s} (|\mathcal{A}| q)^{-1}$.

\subsubsection{MCMC update for $\contact_i$}
Let $\contact_{-i} = (\contact_j)_{j\neq i}$, $A^s_{\contact_i} = \{a \in \mathcal{A}: \contact^s_{a} = \contact_i\}$, and $A_{\contact_i} = (A^s_{\contact_i})_{s \in \jesa}$. We update each $\contact_i$ ($i =1,\ldots, 32$) using a slice sampler for the conditional distribution of
$\contact_i \mid \contact_{-i}, Z, \common, q, A_{\contact_i}$ with density proportional to
\begin{align}
p(\contact_i) & \propto \contact_i^{\sum_{s \in \jesa} |A^s_{\contact_i}|} \prod_{s \in \jesa} \prod_{a \in A^s_{\contact_i}} \frac{1}{ \left(u^s \common_a \contact_i + 1 \right)^{q + C^s_a}}.
\end{align}
Our slice sampler uses the stepping out method as described by \cite{neal2003slice}, with a step width given by 0.01.

\subsubsection{MCMC update for $p^h$ and $p^v$}
Let $p_{-i}^h = (p_j^h)_{j \neq i}$, $p_{-i}^v = (p_j^v)_{j \neq i}$, 
\begin{align}
\Delta^h_{i} = \sum_{s \in \jesa} \sum_{n=1}^{N_s} I(|Z_{n,1}^s - x_{n,1}^s| = i),\\
\Delta^v_{i} = \sum_{s \in \jesa} \sum_{n=1}^{N_s} I(|Z_{n,2}^s - x_{n,2}^s| = i).
\end{align}
We update each $p^h_i$ ($i =1,\ldots, 4$) using a slice sampler for the conditional distribution of $p_i^h \mid p^h_{-i}, Z, x$ with density proportional to
\begin{align}
p(p_i^h) &\propto \exp{\left(\frac{-(p^h_i)^2}{8} \right)}\frac{\prod_{\ell=i}^4\left(\sum_{j=\ell}^4 \exp{(p^h_{j})}/(2 j -1)\right)^{\Delta^h_{\ell}}}{\left(\sum_{j=1}^4 \exp{(p^h_j)}\right)^{\sum_{s \in \jesa} N_s }}.
\end{align}
We update each $p^v_i$ ($i =1,\ldots, 4$) using a slice sampler for the conditional distribution of $p_i^v \mid p^v_{-i}, Z, x$ with density proportional to 
\begin{align}
p(p_i^v) &\propto \exp{\left(\frac{-(p^v_i)^2}{8} \right)}\frac{\prod_{\ell=i}^4\left(\sum_{j=\ell}^4 \exp{(p^v_{j})}/(2 j -1)\right)^{\Delta^v_{\ell}}}{\left(\sum_{j=1}^4 \exp{(p^v_j)}\right)^{\sum_{s \in \jesa} N_s }}.
\end{align}
Our slice samplers use the stepping out method as described by \cite{neal2003slice}, with a step width of 1.

\subsubsection{MCMC update for $\common$}
For $\common$, we depart from slice sampling and instead use elliptical slice sampling \citep{murray2010elliptical}, leveraging the Gaussian prior on $\log(\common)$ to sample from the conditional distribution of $\common \mid \contact, Z, \common, q$ with

Let $\contact_{-i} = (\contact_j)_{j\neq i}$, $A^s_{\contact_i} = \{a \in \mathcal{A}: \contact^s_{a} = \contact_i\}$, and $A_{\contact_i} = (A^s_{\contact_i})_{s \in \jesa}$. We update each $\contact_i$ ($i =1,\ldots, 32$) using a slice sampler for the conditional distribution of
$\contact_i \mid \contact_{-i}, Z, \common, q, A_{\contact_i}$ with density proportional to
\begin{align}
p(\common) & \propto  N(\log(\common), \Sigma) \prod_{s \in \jesa} \prod_{a \in \mathcal{A}} \frac{\common_a^{C^s_a}}{ \left(u^s \common_a \contact^s_a + 1 \right)^{q + C^s_a}}.
\end{align}

\subsection{Importance Sampling Strategy}\label{fullIS}
\subsubsection{Details of Importance Distribution}

The goal of our importance sampling strategy is to approximate the quantity given in (\ref{marglik}). For convenience, we replicate this expression below, swapping in the index $\tau$ instead of $s$ to denote that it is a held-out shoe.
\begin{align*}
p(x^\tau \mid \mathcal{C}^{\tau}, T) &= \mathbb{E}_{\Theta}\left(p(x^\tau| \mathcal{C}^{\tau}, \Theta) \mid (x^s, \mathcal{C}^s)_{s \in T}\right) .
\end{align*}

This expression can be viewed as the composition of two integrals, an outer integral over the posterior distribution of $\Theta$ and an inner integral over the local auxiliary variables $Z^\tau$ and $u^\tau$. 

The outer integral is with respect to the posterior density (\ref{fullposterior}). Running the MCMC strategy outlined in \S \ref{MCMC} with training data $T$ (i.e. $(x^s, \mathcal{C}^s)_{s \in T}$) produces a chain of $L$ draws $(\Theta^{\ell})_{\ell = 1, \ldots, L}$. We can use this chain to approximate the posterior density given in (\ref{fullposterior}), thus approximating the outer integral as
\begin{align*}
p(x^\tau \mid \mathcal{C}^{\tau}, T) &= L^{-1} \sum_{\ell = 1}^L \left(p(x^\tau| \mathcal{C}^{\tau}, \Theta) \mid \Theta^{\ell}.\right).
\end{align*}

For each draw in the Markov chain, we can then approximate the inner integral $p(x^\tau| \mathcal{C}^{\tau}, \Theta)$ in the sum by importance sampling. Recall that the inner integral is given by
\begin{align*}
p(x^\tau| \mathcal{C}^{\tau}, \Theta^{\ell}) &= \int_{0}^{\infty} \frac{u^{N_\tau - 1}}{\Gamma(N_\tau)}\mathbb{E} \left(  \frac{1}{\Gamma(q^{\ell})^{|\mathcal{A}|}}\prod_{a \in \mathcal{A}} \frac{ \Gamma(q^{\ell} + C_a^\tau) \left(\common^{\ell}_{a} (\contact^{\ell})^\tau_{a}\right)^{C^\tau_a}}{\left(u^\tau \common^{\ell}_{a}(\contact^{\ell})^\tau_{a} + 1 \right) ^{q^{\ell} + C^\tau_a}} \right) \text{d}u^\tau\\
&= \int_{0}^{\infty} \sum_{Z^\tau \in \zeta^\tau }  \frac{u^{N_\tau - 1}}{\Gamma(N_\tau)}  \frac{\prod_{n=1}^{N_\tau} \kernel(\Delta_n^\tau)  }{\Gamma(q^{\ell})^{|\mathcal{A}|}}\prod_{a \in \mathcal{A}} \frac{ \Gamma(q^{\ell} + C_a^\tau) \left(\common^{\ell}_{a} (\contact^{\ell})^\tau_{a}\right)^{C^\tau_a}}{\left(u^\tau \common^{\ell}_{a}(\contact^{\ell})^\tau_{a} + 1 \right) ^{q^{\ell} + C^\tau_a}}  \text{d}u^\tau\\
&= \int_{0}^{\infty} \sum_{Z^\tau \in \zeta^\tau } r(u^{\tau}, Z^{\tau}) \text{d}u^\tau.
\end{align*}
Here, $C^\tau_a$ denotes the number of times each $a \in \mathcal{A}$ occurs in $Z^\tau$, $\Delta_n^\tau \in \{-3,3\}^2$ is shorthand for $Z_n^\tau - x_n^\tau$, and 
\begin{align}
\zeta^\tau = \left\{Z^\tau : \Delta_n^\tau \in \{-3,3\}^2, n \in \left\{1,\ldots, N_\tau\right\} \right\}.
\end{align}
The above integral and sum of $r(u^{\tau}, Z^{\tau})$ cannot be evaluated analytically. 

Instead, we use importance sampling to evaluate it for each $\ell \in 1,\ldots, L$, treating the integral and sum of $r(u^{\tau}, Z^{\tau})$ as an expectation of a function of $u^{\tau}, Z^{\tau}$. Implicitly, $r(u^{\tau}, Z^{\tau})$ acts as the product of density and a function of which we are taking the expectation. In practice, there is no need to make a distinction between what serves as the density and what serves as the function --- the product is our target. By drawing values of $Z^{\tau}$ and $u^{\tau}$ from an easy-to-sample-from importance distribution and applying the correct importance weights to our draws, we can target this expectation using a Monte Carlo strategy.

In deriving a good importance distribution, we target three properties: cheaply generated random variates, tractable importance weights, and --- most importantly ---  an importance distribution that serves as a good surrogate for the target expectation. Such a surrogate distributes its density in similar places as target integrand to achieve a low variance estimator. Motivated by these properties, we now derive the importance distributions for $u^{\tau}$ and $Z^{\tau}$.

Before our marginalization of the $\noise$ terms, the distribution of the auxiliary variable $u^{\tau}$ was independent of $Z^{\tau}$, given by 
\begin{align}
u^\tau \sim \text{Gamma}\left( N_\tau, \sum_{a \in \mathcal{A}} \common^{\ell}_{a} (\noise^{\ell})_{a}^{\tau} (\contact^{\ell})^\tau_{a}\right).
\end{align}
Marginalizing $\noise$ introduced additional dependence the two, making their joint distribution unwieldy. For our importance distribution, we opt for independence by replacing each of the $(\noise^{\ell})_{a}^{\tau}$ terms in the original gamma distribution with their expected value $\mathbb{E}((\noise^{\ell})_{a}^{\tau} | q^{\ell}) = q^{\ell}$. This leads to the importance distribution
\begin{align}
u^{\tau} \sim \text{Gamma}\left(N_s,  q^{\ell} \sum_{a \in \mathcal{A}} \common^{\ell}_a (\contact^{\ell})^\tau_{a}\right),
\end{align}
as given in (\ref{importanceu}) within the main text.

For each $Z^{\tau}_n$, we could use the distribution given by (\ref{distz}). However, we can do better. Noting that a factor $\left(\common^{\ell}_{a} (\contact^{\ell})^\tau_{a}\right)$ occurs in (\ref{marginal2}) for each $Z^{\tau}_n = a$, we incorporate these terms in our distribution as well, letting
\begin{align*}
\mathbb{P}(Z^{\tau}_n = x_n^{\tau} + a) & = \frac{\common_{a + x_n^{\tau}} \contact^{\tau}_{a +  x_n^{\tau}} \kernel(a)}{\sum_{b \in B}\common_{b + x_n^{\tau}} \contact^{\tau}_{b +  x_n^{\tau}} \kernel(b)} 
\end{align*}
serve as our importance distribution. This approach is especially helpful when accidentals occur in areas with sparse contact surface. The $\contact$ information in prevents the distribution from overdrawing unlikely gridpoints surrounded by little contact surface.

The resultant full importance distribution has a hybrid density/mass function given by
\begin{align*}
h(u^{\tau}, Z^{\tau}) &= \frac{(u^{\tau})^{N_s - 1}}{\Gamma(N_s)} \frac{\exp{\left(-u^{\tau} q \sum_{a \in \mathcal{A}} \common^{\ell}_a (\phi^{\ell})^{\tau}_a\right)}}{\left( q \sum_{a \in \mathcal{A}} \common^{\ell}_a (\contact^{\ell})^{\tau}_a \right)^{N_\tau}} \prod_{n = 1}^{N_s} \frac{\common^{\tau}_{Z^{\tau}_n} \contact^{\tau}_{Z^{\tau}_n} \kernel(\Delta_n^\tau) }{\sum_{b \in \zeta_n^{\tau}}\common_{b} \contact^{\tau}_{b} \kernel(b - x^\tau_n)}.
\end{align*}

To derive the contribution of each generated sample, we divide the integrand $r(u^{\tau}, Z^{\tau})$ by the importance density $h(u^{\tau}, Z^{\tau})$. The result is
\begin{align}
w(u^{\tau}, Z^{\tau}) &= \frac{r(u^{\tau}, Z^{\tau})}{h(u^{\tau}, Z^{\tau})}\\
&= \frac{\prod_{n=1}^{N_{\tau}} \left( \sum_{b \in \zeta_n^{\tau}}\common^{ell}_{b} (\contact^\ell)^{\tau}_{b} \kernel^\ell(b - x^{\ell}_n) \right)}{ \Gamma(q^\ell)^{|\mathcal{A}|} \left( q^\ell \sum_{a \in \mathcal{A}} \common^{\ell}_a (\contact^\ell)^{\tau}_a \right)^{N_\tau}} \frac{ \exp{(u^\ell q^\ell \sum_{a \in \mathcal{A}} \common^\ell_a (\contact^\ell)^{\tau}_a)}}{ \prod_{a \in \mathcal{A}} \frac{\left(u^\ell \common^\ell_a (\contact^\ell)^\tau_a + 1 \right)^{q^\ell + C^\ell_a}}{\Gamma(C^\ell_a + q^\ell)}}. \label{impweights}
\end{align}

Thus, using one importance sample ($M=1$) for each MCMC draw $\Theta^{\ell} = (\contact^{\ell}, \common^{\ell}, q^{\ell}, (p^h)^{\ell}, (p^v)^{\ell})$ yields the approximation
\begin{align}
p(x^\tau \mid \mathcal{C}^{\tau}, T)  \approx \sum_{\ell=1}^L \frac{\prod_{n=1}^{N_{\tau}} \left( \sum_{b \in B}\common_{b + x_n^{\tau}} (\contact^\ell)^{\tau}_{b +  x_n^{\tau}} \kernel^\ell(b) \right)}{ L \Gamma(q^\ell)^{|\mathcal{A}|} \left( q^\ell \sum_{a \in \mathcal{A}} \common_a (\contact^\ell){\tau}_a \right)^{N_\tau}} \frac{ \exp{(u^\ell q^\ell \sum_{a \in \mathcal{A}} \common^\ell_a (\contact^\ell)^{\tau}_a)}}{ \prod_{a \in \mathcal{A}} \frac{\left(u^\ell \common^\ell_a (\contact^\ell)_a^\tau + 1 \right)^{q^\ell + C^\ell_a}}{\Gamma(C^\ell_a + q^\ell)}}.
\end{align}
where $L$ is the total number of MCMC draws.

\subsubsection{llustration and Discussion Chain Mixing}

\begin{figure}[h]
\centering
 \input{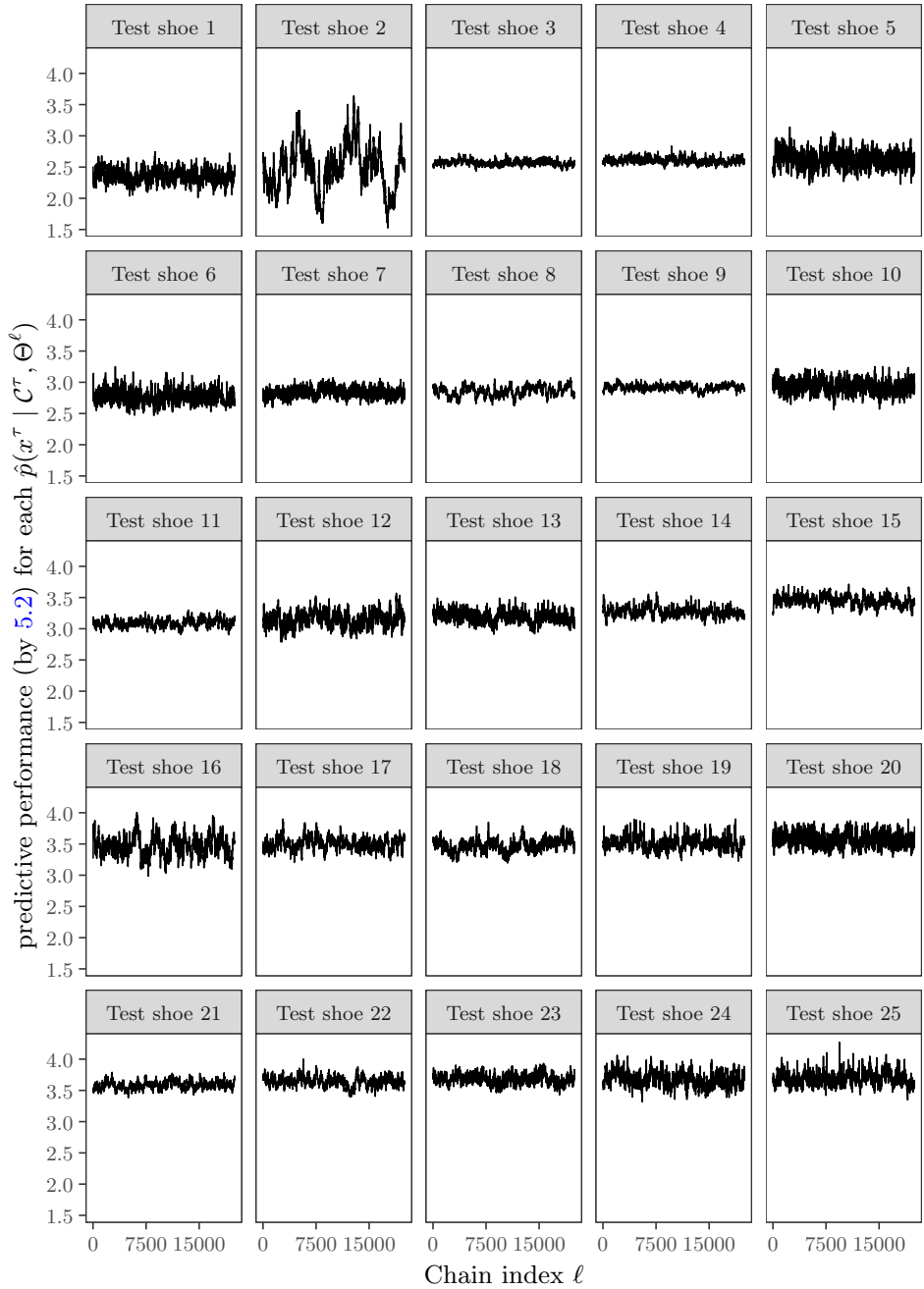} 
\caption{Trace plots demonstrating the Markov chain for the estimated held-out predictive performance per accidental corresponding to shoes 1 through 25 of Data Split 1. The shoes are ordered sequentially by the mean predictive performance, and the results pertain to our full model.} \label{trace1}
\end{figure}

The reliability of the Monte Carlo approximation described in Section~\ref{computation} depends on the mixing of the Markov chain. Here, we demonstrate the mixing of the chain for the targeted quantities by highlighting on the trace plots (Figures~\ref{trace1} and \ref{trace2}) of the held-out probability estimates shown in Figure~\ref{performance} (specifically, those listed for ``Our model'' for Split 1). Such quantities (the posterior probability of held-out data under our fitted model) are the target of our model.

Recall that the quantities shown in Figure~\ref{performance} (and summarized in Table~\ref{resultstab}) were obtained by the following process. The Markov chain strategy outlined in \S\ref{MCMC} was run for 30000 iterations, after which the first 10000 iterations were discarded as warm-up. For each shoe $\tau = 1, \ldots, 50$, the importance sampling estimates
\begin{align}
\hat{p}(x^{\tau}\mid \mathcal{C}^{\tau}, \Theta^{\ell}) &= w(u^{\tau}, Z^{\tau}) 
\end{align}
were computed (by (\ref{impweights})) using one importance sample for each of 20000 remaining chain iterations ($\ell = 1, \ldots, 20000$). The metric given by (\ref{plotmetric}) was then calculated for each $\hat{p}(x^{\tau}\mid \mathcal{C}^{\tau}, \Theta^{\ell})$. To obtain the summaries shown in Figure~\ref{performance}, the mean was taken of each chain. Here, we examine the contents of each of the 50 chains before averaging. 

Figures~\ref{trace1} and \ref{trace2} demonstrate trace plots for each of the 50 held-out shoes over the 20000 iterations of the Markov chain. The chains are presented in order of mean per-shoe performance --- the same sequence they appear on the $x$-axis in Figure~\ref{performance} (Data Split 1). By inspection, almost every chain mixed very well. The notable exception is the chain for held-out shoe 2, having an effective sample size of 19.3 (all others exceed 35). The slower mixing of this case is not necessarily surprising --- this chain appears to be more diffuse than the others, indicating that the probability evaluation is especially uncertain.

 Nonetheless, an effective sample size of 19.3 still provides a reasonable estimate of the mean of the chain. If it is essential to have more accurate estimates of less robust functions (such as more extreme quantiles), the chain would have to be run for a longer number of iterations. 

\begin{figure}[h]
\centering
\input{traceplots2.tex} 
\caption{A companion to Figure~\ref{trace1}. Trace plots demonstrating the Markov chain for the estimated held-out predictive performance per accidental corresponding to shoes 26 through 50 of Data Split 1. The shoes are ordered sequentially by the mean predictive performance, and the results pertain to our full model.} \label{trace2}
\end{figure}

\begin{figure}[h]
\centering
\subfloat[]{\includegraphics[height = 2.1in]{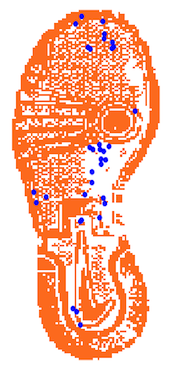}\label{offaccidental1}} \hspace{1cm}
\subfloat[]{\includegraphics[height = 2.1in]{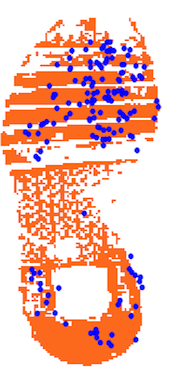}\label{offaccidental2}} 
\caption{The contact surfaces and overlaid accidentals of two example shoes (a) and (b) from the JESA database. In both of these cases, some of the accidentals do not occur on the contact surface}\label{offcontacts}
\end{figure}

\end{document}